\documentclass[structabstract]{aa}
\usepackage{graphicx}
\usepackage{natbib} \bibpunct{(}{)}{;}{a}{}{,}
\usepackage{color}
\usepackage{rotating}
\usepackage{txfonts}

\def\changed{}
 
 \newcommand{\msunpyr}{\,M_\odot\,\mbox{yr}^{-1}}

\newcommand{\kms}{\ifmmode{\,\mbox{km}\,\mbox{s}^{-1}}\else{km/s}\fi}
\newcommand{\msun}{\ifmmode M_{\odot} \else M$_{\odot}$\fi}
\newcommand{\rsun}{\ifmmode R_{\odot} \else R$_{\odot}$\fi}
\newcommand{\lsun}{\ifmmode L_{\odot} \else L$_{\odot}$\fi}
\newcommand{\zsun}{\ifmmode Z_{\odot} \else $Z_{\odot}$\fi}
\newcommand{\velo}{\ifmmode\varv\else$\varv$\fi}
\newcommand{\vinf}{\ifmmode\velo_\infty\else$\velo_\infty$\fi}
\begin{document} 
 
\title{Physics and evolution of the most massive stars in 30\,Dor}
\subtitle{Mass loss, envelope inflation, and a variable upper stellar mass limit}

\titlerunning{Stellar evolution in the upper HRD}
 
\author{G\"otz Gr\"{a}fener\inst{\ref{inst1}}
}
 
\institute{Argelander-Institut f{\"u}r Astronomie der Universit{\"a}t Bonn, Auf dem H{\"u}gel 71, 53121 Bonn, Germany\label{inst1}
}

 
\date{Received ; Accepted}

\abstract{The identification of stellar-mass black-hole mergers with up to $80\,M_\odot$ as powerful sources of gravitational wave radiation led to increased interest in the physics of the most massive stars. The largest sample of possible progenitors of such objects, very massive stars (VMS) with masses up to 300\,$M_\odot$, have been identified in the 30\,Dor star-forming region in the Large Magellanic Cloud (LMC). In this young starburst analogue VMS were found to dominate stellar feedback. Despite their importance the physics and evolution of VMS is highly uncertain, mainly due to their proximity to the Eddington limit.}
{In this work we investigate the two most important effects that are thought to occur near the Eddington limit. Enhanced mass loss through optically thick winds, and the formation of radially inflated stellar envelopes.}
{We compute evolutionary models for VMS at LMC metallicity and perform a population synthesis of the young stellar population in 30\,Dor. We adjust the input physics of our models to match the empirical properties of the single-star population in 30\,Dor as derived in the framework of the VLT-Flames Tarantula Survey (VFTS).}
{Enhanced mass loss and envelope inflation near the Eddington limit have a dominant effect on the evolution of the most massive stars. While the observed mass-loss properties and the associated surface He-enrichment are well described by our new models, the observed O-star mass-loss rates are found to cover a much larger range than theoretically predicted, with particularly low mass-loss rates for the youngest objects. Also, the (rotational) surface enrichment in the O-star regime appears to be not well understood. The positions of the most massive stars in the Hertzsprung-Russell Diagram (HRD) are affected by mass loss and envelope inflation. For instance, the majority of luminous B-supergiants in 30\,Dor, and the lack thereof at the highest luminosities, can be explained through the combination of envelope inflation and mass loss. Finally, we find that the upper limit for the inferred initial stellar masses in the greater 30\,Dor region is significantly lower than in its central cluster R\,136, implying a variable upper limit for the masses of stars.}
{The implementation of mass-loss and envelope physics in stellar evolution models turns out to be essential for the modelling of the observable properties of young stellar populations. While the properties of the most massive stars ($\gtrsim 100\,M_\odot$) are well described by our new models, the slightly less massive O stars investigated in this work show a much more diverse behaviour than previously thought, with potential implications for rotational mixing and angular momentum transport. While the present models are a big step forward in the understanding of stellar evolution in the upper HRD, more work is needed to understand the mechanisms that regulate the mass-loss rates of OB stars and the physics of fast-rotating stars.}
\maketitle

\section{Introduction}
\label{sec:INTRO}

The recent identification of massive black-hole binaries with masses up to $80\,M_\odot$ as powerful sources of gravitational waves \citep{abb2:16,abb1:16} opened the first direct view on the end products of the evolution of the most massive stars. To be able to exploit this new information it is necessary to model stellar populations that are young enough to contain Very Massive Stars (VMS) with masses $\gtrsim$\,$100\,M_\odot$. Because of the steep increase of stellar luminosities with increasing mass, VMS in this regime have luminosities well in excess of $10^6\,L_\odot$. The resulting high $L/M$ ratios $\gtrsim 10^4\,L_\odot/M_\odot$ are bringing such VMS close to the Eddington limit where the radiative acceleration $a_\mathrm{rad}$ equals the gravitational acceleration $g$.

Due to the resulting low effective surface gravities VMS are subject to physical effects affecting their outer envelope structure and stellar winds. The envelopes of VMS are predicted to be subject to the envelope inflation effect, which leads to the formation of radially extended low-density envelopes \citep{ish1:99,pet1:06}. In the limit of low convective efficiency inflated envelopes become unstable above a specific $L/M$ ratio because of a lack of hydrostatic envelope solutions \citep{gra1:12}. This leads to numerical problems in stellar evolution models, which are often tackled by an artificial increase of the convective efficiency \citep[cf.][]{pax1:13}. As a result the inflation effect is suppressed in most evolutionary models. Relaxing this assumption leads to the formation of relatively cool stars with inflated envelopes at the top of the main sequence \citep{gra1:12,koe1:15,san1:15}. 

One key question addressed in this work is whether the existence of the inflation effect can be corroborated by observations. This is particularly interesting because the inflation effect may be related to the still enigmatic group of luminous blue variables (LBVs) and their S-Dor type (radius) variability on timescales of months to years \citep[cf.][]{gra1:12,gra1:20}.

The enhanced mass loss of VMS, and its transition from optically thin O-star winds to optically thick Wolf-Rayet winds have been investigated by numerical wind models \citep{gra1:08,vin1:11} that are supported by systematic empirical studies of samples of VMS \citep{gra1:11,bes1:14}. \citet{gra1:11} pointed out that the underlying reason for the enhanced mass loss of WR stars is their proximity to the Eddington limit, and not their surface chemistry as usually assumed in stellar evolution models. The enhanced mass loss is the reason why VMS often appear as emission-line stars, namely as hydrogen-rich Wolf-Rayet (WNh) stars, even during their main-sequence evolution \citep{ham1:06,sch1:08,sch1:09,cro1:10}. Due to their extreme luminosities the spectral signatures of such very massive WNh stars are detectable in young starbursts even at cosmological distances \citep{wof1:14,gra1:15}. 

The winds of WNh stars are often confused with those of classical Wolf-Rayet (WR) stars. Classical WR stars are massive stars near the end of their evolution, predominantly in the phase of core He-burning. These stars have lost their hydrogen-rich envelopes during their previous evolution and have much lower masses. They reach high Eddington factors due to the high luminosity of their He-burning cores. Classical WR stars have different physical and wind properties than the more massive core hydrogen-burning WNh stars, but they are usually treated with the same mass-loss prescriptions in evolutionary models.

In this work we construct evolutionary models that take the enhanced mass-loss rates and the envelope inflation of VMS into account. We compute model grids for massive main-sequence stars up to $500\,M_\odot$ and compare the results with the so far best studied sample of VMS obtained in the VLT-Flames Tarantula Survey \citep[VFTS;][]{eva1:11}. The VFTS is a near-complete spectroscopic survey of $\approx$\,$1000$ hot luminous stars in 30\,Dor in the LMC, the closest analogue of a young starburst in our neighbourhood, with a well-controlled brightness cut-off. All single stars in the VFTS sample have been analysed individually, and their star-formation history (SFH) and rotational-velocity distribution have been investigated in detail \citep{sch2:18,sch4:18,ram1:13,duf1:13}. This makes it an ideal target for detailed population synthesis modelling, enabling quantitative predictions of the expected numbers of stars in specific evolutionary phases.

In the following we describe our evolutionary models in Sect.\,\ref{sec:MESA}, and the population synthesis in Sect.\,\ref{sec:POP}. In Sect.\,\ref{sec:30DOR} we compare our models with the observed population of VMS in 30\,Dor, and in Sect.\,\ref{sec:DISC} we discuss the results. The most important conclusions are summarised in Sect.\,\ref{sec:CON}.

\section{Stellar evolution models}
\label{sec:MESA}

We use the Modules for Experiments in Stellar Astrophysics (MESA) version {\changed 11701} \citep{pax1:11,pax1:13,pax1:15,pax1:18,pax1:19} to compute single-star evolutionary models covering the main-sequence phase from zero to 99\% hydrogen exhaustion in the stellar core. The same input physics as in \citet{sch1:18,sch1:19} are used, except for the treatment of convection in the outer stellar envelope and stellar-wind mass loss (see below). In particular, we adopt a wind-scaling as a function of $X_{\rm Fe}/X_{{\rm Fe}, \odot}$, as described by \citeauthor{sch1:19} We use the MESA default chemical composition from \citet{gre1:98} scaled to an initial metallicity of $Z=0.008$ (compared to the corresponding solar metallicity of $Z_\odot = 0.0169$) to match the metallicity, and in particular $X_{\rm Fe}$, in the LMC.

Because of our interest in the envelope inflation effect we employ standard mixing-length theory (MLT) instead of the modified treatment of superadiabatic convection in radiation-dominated regions \citep[MLT++,][]{pax1:13} that is used by default in MESA models. Furthermore, we introduce a new treatment of stellar-wind mass loss to meet the requirements for VMS near the Eddington limit. In the following sections these new mass-loss relations are described.

\subsection{The winds of OB stars}
\label{sec:OB}

The most widely used mass-loss prescription for OB stars is the relation by
\citet{vin1:00,vin1:01}. These theoretical mass-loss estimates are based on a
Monte-Carlo approach that makes use of the first-order Sobolev
approximation. In the Sobolev approximation spectral lines are assumed to be
infinitely narrow, but their flux-mean opacity is high due to the effective
line broadening through Doppler shifts in the wind. Consequently, the
radiative acceleration depends predominantly on the velocity gradient
$\partial\varv/\partial r$, leading to the same wind physics as in \citet[][CAK]{cas1:75} who described the first closed theory for the formation of optically thin OB-star winds.

Notably, the Sobolev approximation does not apply in the limit of large wind
optical depths where the photon mean-free path becomes small, and Doppler
shifts become inefficient \citep[cf.\ the discussion in][]{gra1:17}. In this
work we therefore only apply the \citeauthor{vin1:01} relation in the
optically thin regime. Furthermore, we change the absolute value and the slope
of $\dot{M}$ as a function of luminosity $L$, to match the average observed mass-loss
rates of O\,stars in 30\,Dor as in \citet[][]{bes1:14}.

For the absolute calibration of our mass-loss relations with empirical results in Sect.\,\ref{sec:MDOT30} we adopt a wind clumping factor of $D=10$. This means that we assume that the wind consists of dense clumps with a density that is a factor $D$ higher than the mean density, while the inter-clump medium is void. Clumping constitutes the main uncertainty in the empirical mass-loss rates of OB and WR stars. It increases the line strengths of recombination lines, and therefore reduces the empirical mass-loss rates derived from this type of lines by a factor $\sqrt{D}$ \citep{hil1:91,ham1:98}.

For the most massive O and WNh stars in 30\,Dor \citet{bes1:14} found clumping
factors of the order of $D=10$, based on the relative strengths of the
electron-scattering wings of strong recombination lines
\citep[cf.\ also][]{vin1:12}. For O stars with weaker winds even higher
clumping factors up to $D \approx 100$ have been discussed, based on the
strength of non-saturated UV resonance lines \citep{ful1:06}. However, taking into account
the influence of optically thick clumps on the UV line profiles
\citet{osk1:07,sun1:14} derived more moderate values of the order of 10. Here
we adopt $D=10$ for all stars in our sample, resulting in a reduction of the
empirical mass-loss rates by a factor 3.3 compared to un-clumped winds.

Our comparison with observations in Sect.\,\ref{sec:MDOT30} leads to a
considerable reduction of O-star mass-loss rates, and a shallower slope as a
function of $L$. For the slope we adopt the value of 1.45 derived by
\citet{bes1:14} and replace it in Eq.\,24 of \citet{vin1:01} instead of the
original value of 2.194. This substitution already lowers the mass-loss rates in the VFTS O-star range
considerably. To match the observations, we further apply a constant factor of
0.8 to the altered mass-loss relation. For the dependence of $\dot{M}$ on the
rotational velocity we use the standard relation implemented in MESA
\citep[Eq.\,26 in][]{pax1:13}

\subsection{The winds of WNh stars}
\label{sec:GAMMA}

The most massive and luminous main-sequence stars are thought to appear spectroscopically as WNh stars \citep{gra1:08,cro1:10}. WNh stars are therefore the extension of the O-star regime towards higher masses and luminosities. Previous implementations of wind mass loss in stellar evolution models do not take the transition from the optically thin winds of OB stars to the optically thick winds of WNh stars into account. Instead, they extrapolate the \citeauthor{vin1:00} relations to higher luminosities. {\changed Only when the surface composition becomes substantially deficient in hydrogen,} the mass-loss relations switch to Wolf-Rayet type mass loss, where no distinction between the mass-loss of very massive WNh stars, and the mass loss of classical core He-burning WR stars is made. {\changed For instance, in the default 'DUTCH' MESA wind-scheme the empirical mass-loss relation from \citet{nug1:00} is used for hot stars ($T_{\rm eff} > 10$\,kK) with hydrogen surface mass fractions $X_{\rm s} < 0.4$, otherwise the \citeauthor{vin1:00} relations are used.}

In this work we follow previous theoretical \citep{gra1:08,vin1:11} and empirical \citep{gra1:11,bes1:14} studies of the mass loss properties of VMS that suggest that the proximity to the Eddington limit determines the formation of optically thick WR-type winds. According to these studies WR-type winds show  a distinct, strong dependence on the proximity to the Eddington limit, which is most easily expressed via a dependence on the classical Eddington factor
\begin{equation}
 \Gamma_{\rm e}=\frac{\chi_{\rm e}L}{4\pi c GM}
  = 10^{-4.813} \times \left( 1 + X_{\rm s} \right) \frac{L/L_\odot}{M/M_\odot}.
\label{eq:GAMMA_E}
\end{equation} 
Here we define $\Gamma_{\rm e}$ as the ratio between the radiative acceleration on free electrons (with the electron-scattering opacity $\chi_{\rm e}$ for a fully ionised plasma) and the gravitational acceleration of the star. Because $\chi_{\rm e}$ depends only on the number of free electrons per gram, $\Gamma_{\rm e}$ depends only on fundamental stellar parameters, namely the hydrogen surface abundance $X_{\rm s}$ and the $L/M$ ratio. In this picture the mass loss of VMS, which are still in the core hydrogen burning phase, differs from the mass loss of classical core He-burning Wolf-Rayet stars because of their different $L/M$ ratios, effective temperatures, and chemical compositions.

In the following we concentrate on VMS at LMC metallicity. We employ the empirical mass-loss relation for VMS in 30\,Dor from \citet{bes1:14} who derived a relation of the form
\begin{equation}
\label{eq:BEST}
\log(\dot{M}_{\rm thick}) = 5.22 \times \Gamma_{\rm eff} - 0.5 \times \log(D)
  -2.6,
\end{equation}
for the optically-thick winds of WNh stars. Here $D$ denotes the wind clumping factor (cf.\ Sect.\,\ref{sec:MDOT30}), and $\Gamma_{\rm eff}$ the effective Eddington factor including centrifugal acceleration \citep[cf.\ the discussions in][]{gra1:08,gra1:15}. As $\Gamma_{\rm eff}$ varies over the stellar surface we use an intermediate value which lies between the value at the poles, where the centrifugal acceleration is zero, and the equator where $g_{\rm rot} = \Omega^2R$
\begin{equation}
 \Gamma_{\rm eff}= \Gamma_{\rm e} + 0.5 \times \Gamma_{\rm rot}
  \equiv \Gamma_{\rm e} + 0.5 \times \frac{\Omega^2R^3}{GM}.
\end{equation}
{\changed As a result, our mass-loss prescription is sensitive to a combination of the proximity to the Eddington limit and the critical rotation rate similar to the $\Gamma\Omega$-limit discussed by \citep{mae1:00,lan1:97}.}

As a criterion to switch between the regimes of WR-type winds and optically thin OB-star winds we evaluate their sonic-point optical depth $\tau_{\rm s}$. This approach is motivated by the theory of optically thick winds, where the properties of winds with large $\tau_{\rm s}$ are largely constrained by the conditions at the sonic point and $\tau_{\rm s}$ itself \citep{nug1:02,gra1:17}. As described in Sect.\,\ref{sec:OB}, the different wind physics for the optically thin winds of OB stars (with small values of $\tau_{\rm s}$) leads to lower mass-loss rates.

To decide whether a stellar model with given surface values of $M, L, R, \Omega$ and $X_{\rm s}$ can develop a WR-type wind we estimate the value of $\tau_{\rm s}$ that its wind {\em would have} if it had an optically thick wind. To this purpose we compute $\dot{M}_{\rm thick}$ from Eq.\,\ref{eq:BEST} and estimate the terminal wind velocity $\varv_\infty$ using the empirical relation from \citep[][]{lam1:95} with $\varv_\infty/\varv_{\rm esc}^{\rm eff} = 2.51$ \citep[cf.\ also][]{bes1:14}. For given $\dot{M}$ and $\varv_\infty$ we estimate $\tau_{\rm s}$ using a relation between $\tau_{\rm s}$ and the wind efficiency factor $\eta \equiv \dot{M}\varv_\infty / (L/c)$ from \citet{gra1:17}. $\eta$ denotes the ratio between the mechanical wind momentum and the momentum of the radiation field driving the wind. It is related to $\tau_{\rm s}$ via the number of absorption and re-emission processes needed to drive the wind. \citet{gra1:17} derived an approximate relation of the form
\begin{equation}
\label{eq:TAUS}
    \tau_{\rm s} \approx \frac{\dot{M}\varv_\infty}{L/c}\left(1-\frac{\varv_{\rm esc}^2}{\varv_\infty^2}\right).
\end{equation}

In our models we use $\tau_{\rm s}$ from Eq.\,\ref{eq:TAUS} as a criterion for the onset of WR-type mass loss. For cases where $\tau_{\rm s} < \tau_1$ we use the mass-loss prescriptions for optically thin winds from Sect.\,\ref{sec:OB}, and for $\tau_{\rm s} > \tau_2$ we use the $\Gamma$-dependent mass-loss relation Eq.\,\ref{eq:BEST}. For $\tau_1 < \tau_{\rm s} < \tau_2$ we interpolate between both cases. As demonstrated in Sect.\,\ref{sec:POP} we can reproduce the observed mass-loss rates for VMS in 30\,Dor from \citet{bes1:14} very well with values of $\tau_1=2/3$ and $\tau_2=1$.

\section{Population synthesis models}
\label{sec:POP}

In this section we describe the population synthesis models that we use to compare the models from Sect.\,\ref{sec:MESA} with observations. 

\subsection{Stellar evolution models}
\label{sec:STD}

Each of our population synthesis models is based on a set of stellar evolution models for massive stars with initial masses in the range of 9-500\,$M_\odot$ and initial rotational velocities between 0 and 90\,\% of the critical rotation rate.  Typically, each model grid consists of models with 18 different masses and 11 different rotational velocities, that is, a total of 198 models. As start models we use rotating, chemically homogeneous models in thermal equilibrium, with a default initial composition with $Z = 0.008$ (cf.\ Sect.\,\ref{sec:MESA}). The models cover the main-sequence phase up to 99\% hydrogen exhaustion.

\subsection{Population synthesis}

The goal of our population synthesis models is to compute the fraction of stars in a specific stellar population that show one or more observables $o$ in a given parameter range $o_i$. The population is defined through a given initial mass function $\xi(M) = \frac{{\rm d}N}{{\rm d}M}$, initial rotational velocity distribution $P(\varv_{\rm ini})=\frac{{\rm d}N}{{\rm d}\varv_{\rm ini}}$, the star-formation history  $S(t)=\frac{{\rm d}N}{{\rm d}t}$, and the total number of stars $N$.

Our models provide stellar parameters $p$ as a function of time $t$ for each evolutionary track with a given initial mass $M$ and initial fraction of the critical rotation rate $\omega \equiv \Omega/\Omega_{\rm crit}$. We further determine the main-sequence lifetime $T_{\rm MS}$ for each evolutionary track and define the relative age $t_{\rm r} \equiv t/T_{\rm MS}$, so that the $p$ are given as a function of the form $p(t_{\rm r}, M, \omega)$ on the model grid. To compute the $o_i$ we transform the $p(t_{\rm r}, M, \omega)$ onto a homogeneous grid that is logarithmic in $M$ and linear in $\omega$ and $t_{\rm r}$, through linear interpolation between evolutionary tracks. In this step we typically use 100 grid points in $M$ and $t_{\rm r}$, and 64 grid points in $\omega$, where $t_{\rm r}$ lies between 0 and 1, $M$ between 9 and $500\,M_\odot$, and $\omega$ between 0 and 0.9. Based on the $p(t_{\rm r}, M, \omega)$ we then compute any observable $o(t_{\rm r}, M, \omega)$ of interest on the same grid.

To compute the probability distribution of the different $o_i$ for the adopted stellar population we need to compute a weight function $W(t_{\rm r}, M, \omega)$ that takes the star formation history $S(t)$ and the initial distributions of $M$ and $\omega$ into account. The probability $P_i$ to find an observable $o$ in a given parameter range $o_i$ is then
\[
P_i = \frac{ \sum_{t_{\rm r}, M, \omega} W(t_{\rm r}, M, \omega) \left|_{o(t_{\rm r}, M, \omega) \in o_i}\right. }
{ \sum_{t_{\rm r}, M, \omega} W(t_{\rm r}, M, \omega) },
\]
where the summation is a sum over all grid points, and the weight $W(t_{\rm r}, M, \omega)$ takes the initial distribution of $M$ and $\omega$ and the sampling of $t_{\rm r}$, $M$ and $\omega$ into account. We write  $W(t_{\rm r}, M, \omega)$ in the form
\[
W(t_{\rm r}, M, \omega) = W_t(t_{\rm r}) \times W_M(M) \times W_\omega(\omega).
\]
and compute $W_t$, $W_M$ and $W_\omega$ in the following way.

As we use linear time steps $\Delta t_r$, $W_t$ is given by the product of the star formation history $S(t)$ and the derivative ${\rm d}t/{\rm d}t_r=T_{\rm MS}$
\[
W_t = S(t) \times T_{\rm MS}(M,\omega),
\]
with $t=t_{\rm r} \times T_{\rm MS}(M, \omega)$.

For $M$ we use a logarithmic grid with ${\rm d}M/{\rm d}\log(M) = M$, and adopt 
an initial mass function $\xi(M) \propto M^{-\gamma}$, so that
\[
W_M = \xi(M) \times \frac{{\rm d}M}{{\rm d}\log(M)} \propto M^{1-\gamma}.
\]

For $\omega$ we use a linear grid, but the initial rotational distribution $P(\varv_{\rm ini})$ is given as a function of the rotational velocity, so that
\[
W_\omega = P(\varv_{\rm ini}) \times \frac{{\rm d} \varv_{\rm ini}}{{\rm d} \omega}.
\]
Here we compute ${{\rm d} \varv_{\rm ini}}/{{\rm d} \omega}$ numerically, where we define $\varv_{\rm ini}(M,\omega) \equiv \varv_{\rm rot}(t_r=0.01, M, \omega)$ to avoid problems due to the rapid adjustment phase of the internal rotation profile at the start of each evolutionary sequence.

\section{Very massive stars in 30\,Dor}
\label{sec:30DOR}

In this section we compare the results of our population synthesis with the empirically derived properties of the stellar population in 30\,Dor. The empirical data consist of the sub-sample of single main-sequence stars from the VLT-Flames Tarantula Survey \citep[VFTS;][]{eva1:11} as compiled by \citet{sch2:18}. In particular, the empirical data are based on detailed spectroscopic analyses of the most massive stars in 30\,Dor from \citet{bes1:14}, O-type giants and supergiants from \citet{ram1:17}, O-type dwarfs from \citet{sab1:17}, B-type supergiants from \citet{mce1:15} and B-dwarfs from \citet{sch2:18}.

For the population synthesis we used the star-formation history $S(t)$ and initial mass function (IMF) $\xi(M)$ for 30\,Dor from \citet{sch2:18}, together with the rotational velocity distribution $P(\varv)$ for single O stars from \citet{ram1:13}. Analogous to the work of \citeauthor{sch2:18} the present work thus focuses on single stars with a luminosity $\gtrsim$\,$10^{4.5} L_\odot$, well above the VFTS brightness cut-off. This means that only stars with masses $\gtrsim$\,15\,$M_\odot$ and ages $\lesssim$\,12\,Myr are considered in the present analysis.

We further employed evolutionary models for single stars as described in Sect.\,\ref{sec:MESA}, tailored to match the empirical properties of the massive single-star population in 30\,Dor. Because of our focus on the physics of very massive stars these models are designated as VMS models in the remainder of this work. We compare these models with empirical data from VFTS and models based on alternative input physics and/or common standard assumptions.

\subsection{Mass distribution}

\label{sec:NvsL}

\begin{figure}
    \centering
    \includegraphics[scale=0.53]{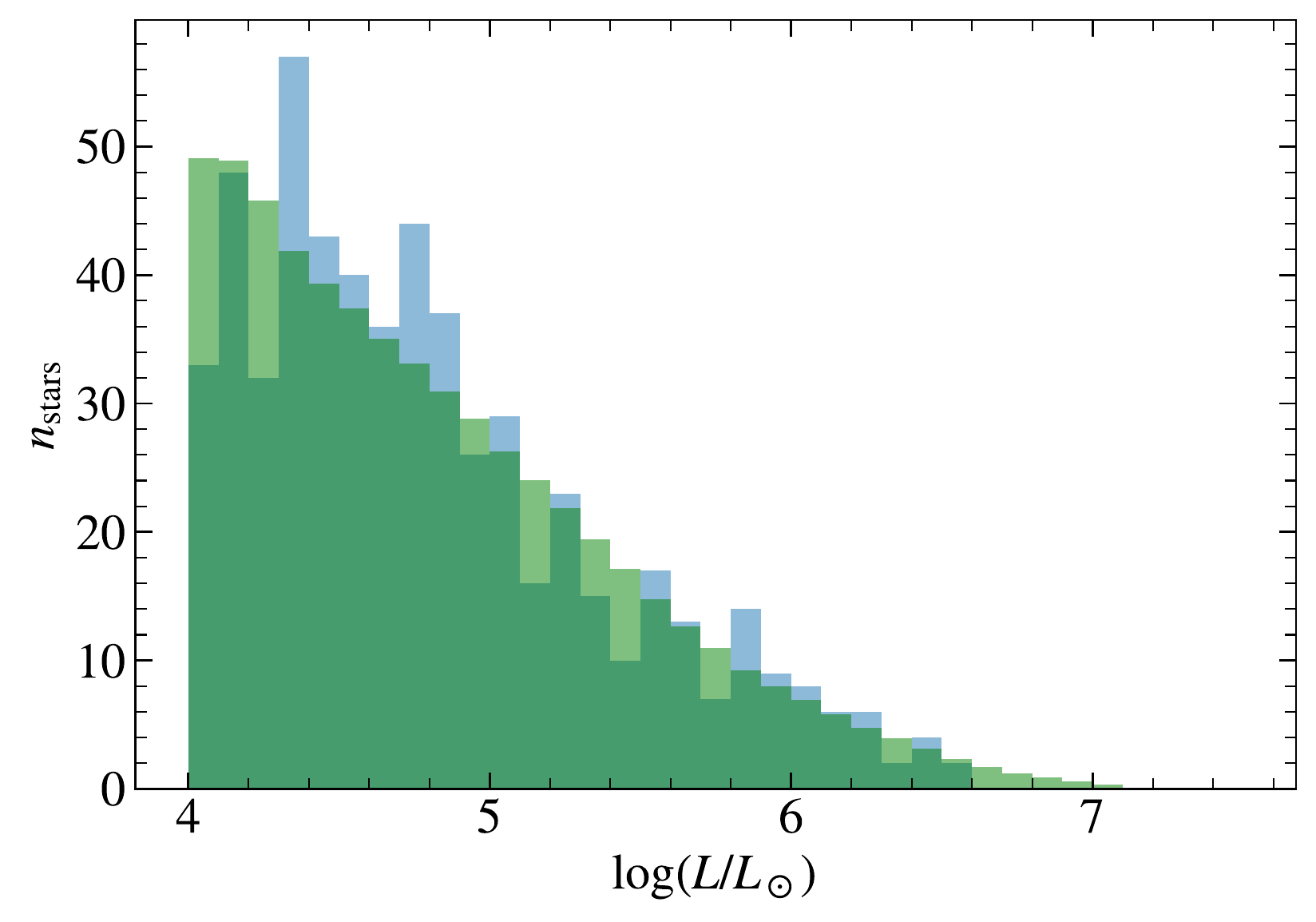}
    \caption{{\changed Observed luminosity distribution of single stars in 30\,Dor compared to theoretical predictions. The histogram shows} the observed numbers of single stars in the VFTS sample as function of luminosity (blue) compared to the theoretical distribution resulting from our VMS population synthesis {\changed models} described in Sect.\,\ref{sec:30DOR} (green). Both distributions are plotted transparent so that overlaps appear in dark green.}
    \label{fig:NvsL}
\end{figure}

An important input parameter for our population synthesis models is the exponent $\gamma$ of the IMF with $\xi(M)\propto M^{-\gamma}$. In Fig.\,\ref{fig:NvsL} we compare the present-day luminosity distribution resulting from our population synthesis with the observed distribution in the VFTS sample. In this plot we normalised the synthetic distribution to match the total number of 288 observed, putatively single stars above $\log(L/L_\odot) = 4.5$. We further adjusted $\gamma$ to match the number of 20 observed VMS in the range $6.0 < \log(L/L_\odot) < 6.6$ (roughly corresponding to a mass range of 70--200\,$M_\odot$). For our VMS models this was achieved using a value of $\gamma=1.977$, slightly higher than the value of $\gamma=1.90_{-0.26}^{+0.37}$ derived by \citet{sch2:18}. Using alternative models employing the {\changed DUTCH MESA wind scheme (cf.\ Sect\,\ref{sec:GAMMA})} we obtain $\gamma = 1.935$, in agreement with \citeauthor{sch2:18}\ Altogether our models confirm a top-heavy IMF in comparison to the standard \citet{sal1:55} IMF with $\gamma=2.35$, but with a moderate impact of the adopted mass-loss prescription on the present stellar masses and therefore on $\gamma$. In the remainder of this work we use $\gamma=1.977$ to achieve consistency between our models and observations.

Notably, there are no stars with luminosities higher than $\log(L/L_\odot) =6.6$ in the VFTS sample while our VMS models predict 4.7 stars above this luminosity. This upper luminosity limit corresponds to an upper initial-mass limit of $M_{\rm ini}\approx 200\,M_\odot$. We will discuss the relevance of this upper mass limit in Sect.\,\ref{sec:ulim}. For the investigation of the physical properties of VMS in the remainder of this work, we consider the consistency between observed and predicted stellar numbers below $200\,M_\odot$ as satisfactory.

\subsection{Mass loss rates}
\label{sec:MDOT30}

\begin{figure}[tbp!]
    \centering
    \includegraphics[scale=0.45]{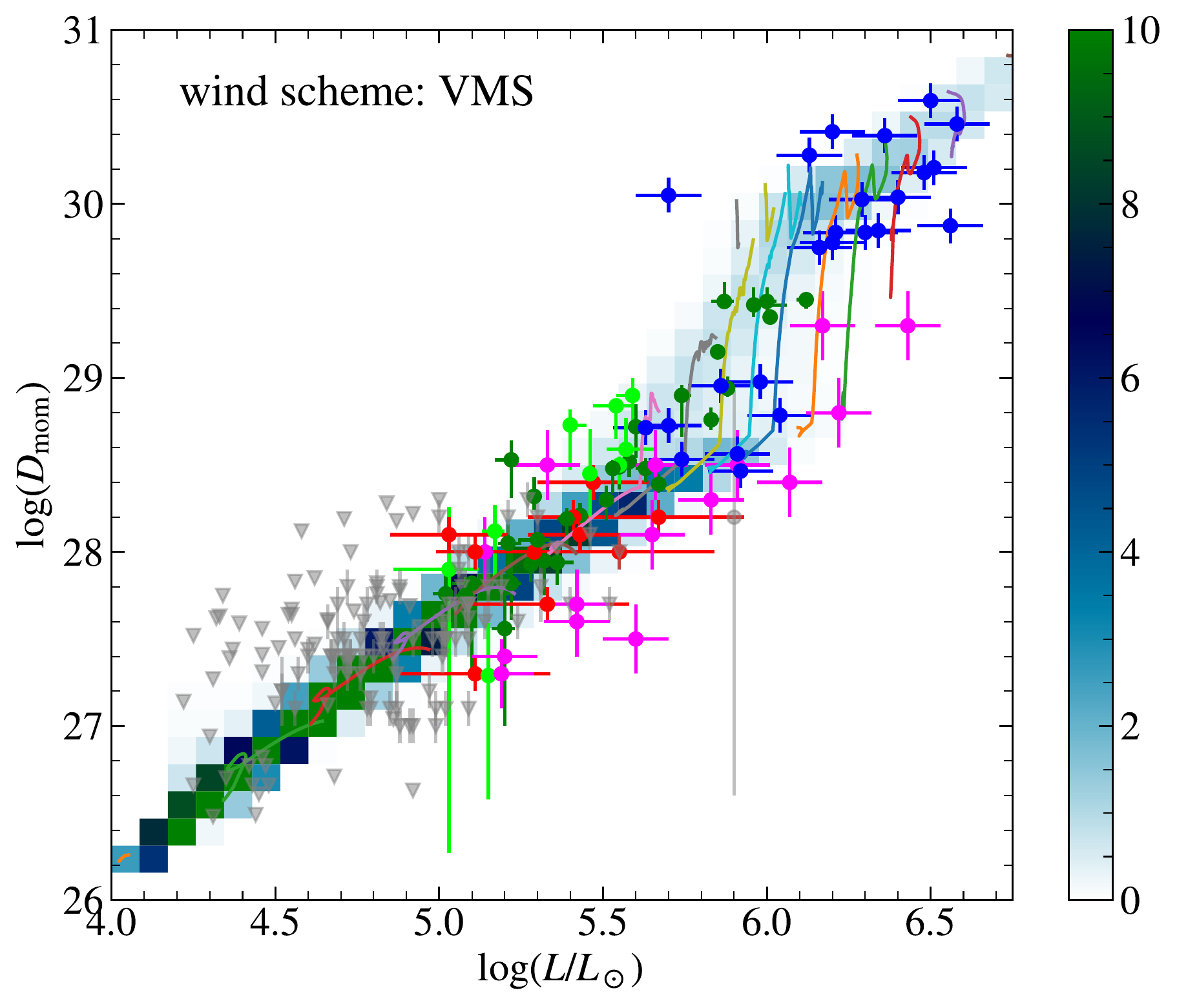}
    \caption{Wind-momentum -- luminosity relation for O and WNh stars in 30\,Dor. Different symbols indicate empirical results for the most massive stars in 30\,Dor from \citet{bes1:14} in blue, O-type giants and supergiants from \citet{ram1:17} Tables C.4 (dark green) and C.5 (light green), and O-type dwarfs from \citet{sab1:17} Tables A.1 (red) and A.2 (pink). Upper limits are indicated in grey. Evolutionary tracks are indicated by coloured lines analogous to Fig.\,\ref{fig:HRD} but restricted to the O\,star range. Predicted absolute numbers of stars according to our population synthesis models are mapped and colour coded as indicated by the colour bar on the right.}
    \label{fig:DMOM_VMS}
\end{figure}

\begin{figure}[tbp!]
    \centering
    \includegraphics[scale=0.45]{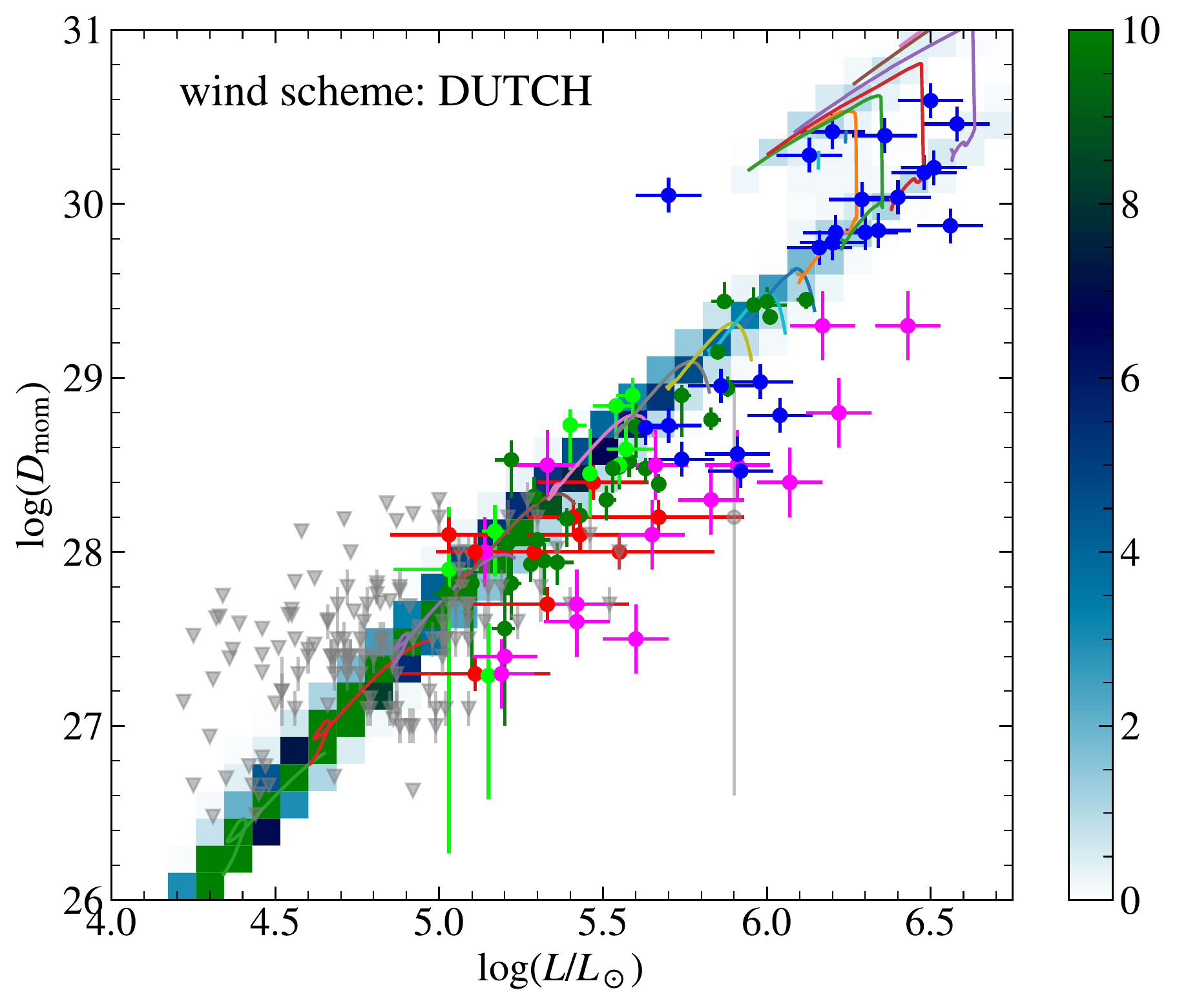}
    \caption{Wind-momentum -- luminosity relation as in Fig.\,\ref{fig:DMOM_VMS} compared with population synthesis models based on standard MESA models using the DUTCH wind scheme.}
    \label{fig:DMOM_DUTCH}
\end{figure}

\begin{figure}[tbp!]
    \centering
    \includegraphics[scale=0.45]{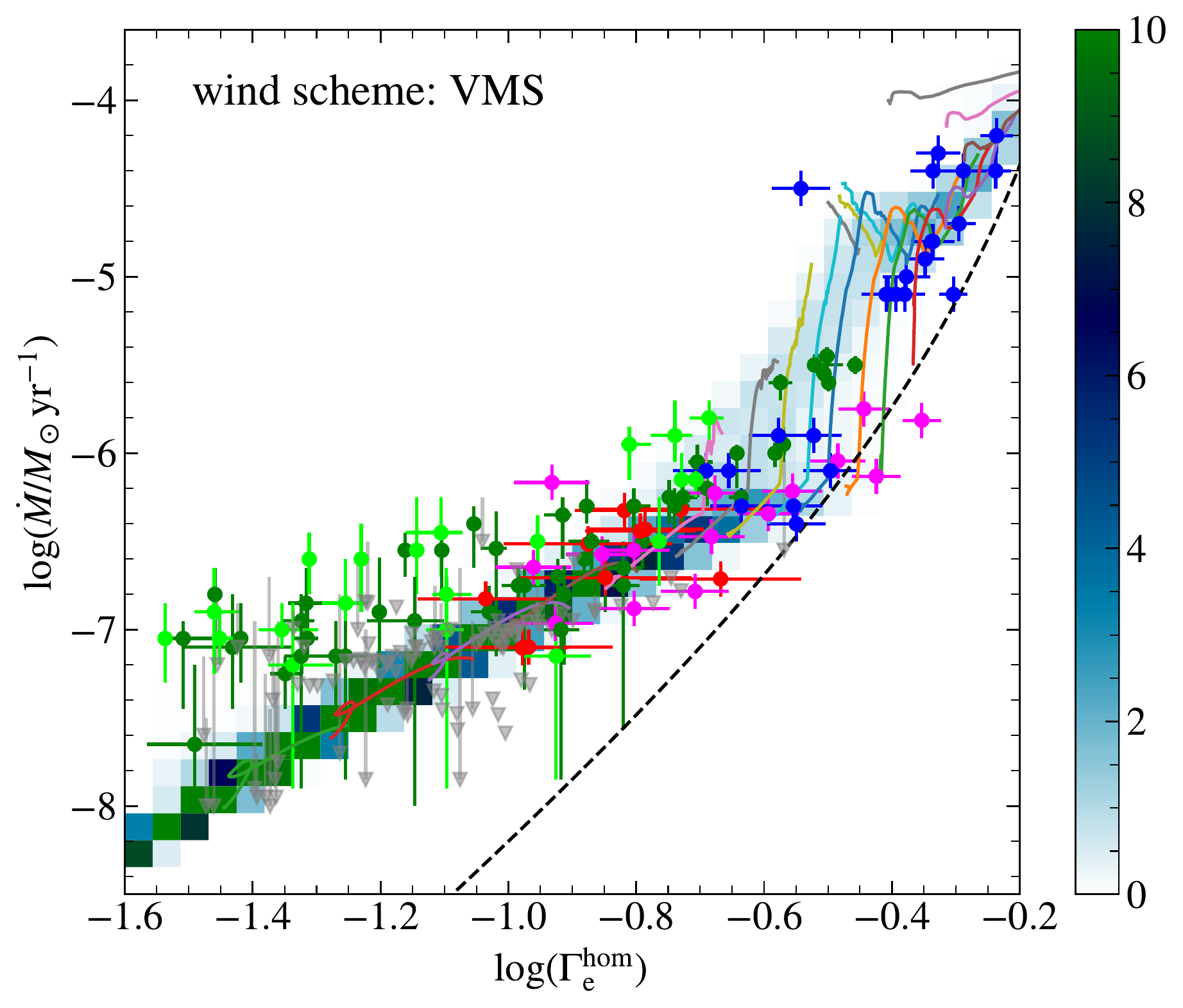}
    \caption{{\changed Mass loss as function of Eddington factor.}  Mass-loss rates $\dot{M}$ as a function of the estimated classical Eddington factor for chemically homogeneous stars $\Gamma_{\rm e}^{\rm hom}$ (see text) analogous to Fig.\,\ref{fig:DMOM_VMS}. The  $\dot{M}(\Gamma_{\rm e})$ relation for R\,136 from \citet{bes1:20} is indicated by the dashed curve.}
    \label{fig:GAMMA_VMS}
\end{figure}

To test the validity of the mass-loss prescriptions employed in our models we compare the mass-loss relations resulting from our population synthesis models directly with empirical data. In Figs.\,\ref{fig:DMOM_VMS} and \ref{fig:DMOM_DUTCH} the resulting modified wind momenta $D_{\rm mom} = \dot{M}\varv_\infty\sqrt{R/R_\odot}$ are shown as a function of luminosity $L$. According to the CAK theory for optically thin winds, OB stars are expected to follow a tight wind-momentum luminosity relation \citep[WLR;][]{kud1:99}, largely independent of secondary parameters such as mass or radius.

In Figs.\,\ref{fig:DMOM_VMS} and \ref{fig:DMOM_DUTCH} we indicate the expected numbers of stars in this diagram according to our population synthesis models by colour maps. The synthetic $D_{\rm mom}$ were computed based on mass-loss rates $\dot{M}$, radii $R$ and stellar masses $M$ from the underlying stellar evolution models. The $\varv_\infty$ were computed from the resulting effective escape velocities using a ratio of $\varv_\infty/\varv_{\rm esc}^{\rm eff} = 2.51$ as described in Sect.\,\ref{sec:GAMMA}. The empirical wind momenta were compiled from \citet{ram1:17,sab1:17,bes1:14} and cover the complete sample of single O and WNh stars in the VFTS, i.e., they include all apparently single main-sequence stars with effective temperatures $> 29$\,kK. The same temperature cut-off has been applied to the theoretical results.

Our VMS models in Fig.\,\ref{fig:DMOM_VMS} show a good agreement with observations. In particular, the location of the transition from O-star winds to the enhanced mass-loss in the WNh phase is reproduced well. On the other hand, the WLR resulting from the oft-used DUTCH wind scheme in Fig.\,\ref{fig:DMOM_DUTCH} matches the WNh stars well, but over-estimates the mass-loss rates of the O stars in the VFTS sample. Also, the slope of the WLR appears to be too steep in the O-star range. For both wind schemes it is noticeable that the synthetic WLRs are much narrower than the observed one. In particular for O stars the observed WLR shows a large spread in $D_{\rm mom}$, where dwarfs tend to have lower wind momenta than giants and supergiants by up to a factor 10. As we will show later, this may hint on substantially reduced mass-loss rates for younger O stars.

A novelty in our VMS models is the treatment of the transition between the regimes of optically thin O-star winds and optically thick WNh winds. This transition occurs qualitatively differently in the VMS and DUTCH approaches. In the VMS wind scheme in Fig.\,\ref{fig:DMOM_VMS} stars with luminosities above $\log(L/L_\odot)\approx 5.5$ reach high enough Eddington factors at some point during their main-sequence evolution to perform a quick transition to enhanced WR-type mass-loss rates. In the WNh regime the evolution then proceeds at nearly constant luminosity (cf.\ Fig.\,\ref{fig:HRD}). On the other hand, the DUTCH models in Fig.\,\ref{fig:DMOM_DUTCH} show no marked transition between the two regimes but change their behaviour at some point during the WNh stage, when they reach a helium surface mass fraction of $Y_{\rm s}=0.4$. Before this point they evolve continuously with increasing luminosity from the O-star regime into the WNh stage. After they reached $Y_{\rm s}=0.4$ their mass loss rates increase, and their evolution continues with decreasing luminosity.

As noted earlier, the absolute mass-loss rates in the WNh range are reproduced well by both mass-loss prescriptions. One notable outlier in Figs.\,\ref{fig:DMOM_VMS} and \ref{fig:DMOM_DUTCH} is VFTS\,108, which has a WR-type mass-loss rate despite of its low luminosity of $\log(L/L_\odot)=5.7$. This star is highly enriched in helium ($Y_{\rm s}\approx0.8$) and most likely already in the core helium-burning phase. \citet{bes1:14} demonstrated that this star matches the adopted $\dot{M}(\Gamma_{\rm eff})$ relation (Eq.\,\ref{eq:BEST}) precisely, if a lower mass, corresponding to the core helium-burning case, is adopted.

In Fig.\,\ref{fig:GAMMA_VMS} we put the $\Gamma_{\rm e}$-dependent mass-loss prescription (Eq.\,\ref{eq:BEST}) in our VMS models to a direct test. As $\Gamma_{\rm e}$ is not directly observable, we estimate the Eddington factor $\Gamma_{\rm e}^{\rm hom}(L, X_{\rm s})$ for the chemically-homogeneous case instead, where we compute $\Gamma_{\rm e}^{\rm hom}(L, X_{\rm s})$ from Eq.\,\ref{eq:GAMMA_E} using the mass estimate $M_{\rm hom}(L, X_{\rm s})$ for a chemically-homogeneous star from \citet[][]{gra1:11}. Notably, $\Gamma_{\rm e}^{\rm hom}(L, X_{\rm s})$ depends only on observable quantities. In reality, most stars are of course not chemically homogeneous, but have a chemical profile with an increasing mean molecular weight towards their centre. For fixed values of $M$ and $X_{\rm s}$ this results in a higher luminosity than for a chemically homogeneous star. The other way round, for given $L$ and $X_{\rm s}$, $M_{\rm hom}(L, X_{\rm s})$ constitutes a strict upper limit to the real mass of a star, and $\Gamma_{\rm e}^{\rm hom}$ a lower limit to $\Gamma_{\rm e}$.

In Fig.\,\ref{fig:GAMMA_VMS} we plot the synthetic and observed mass-loss rates $\dot{M}$ for our O/WNh sample as a function of  $\Gamma_{\rm e}^{\rm hom}$. To allow for a direct comparison between theory and observations, the $\Gamma_{\rm e}^{\rm hom}(L, X_{\rm s})$ are computed in an identical manner for models and observations. Analogous to the WLR in Fig.\,\ref{fig:DMOM_VMS} the theoretical mass-loss rates increase rapidly as soon as the formation of optically thick winds becomes possible. The transition between O and WR-type winds happens at $\log(\Gamma_{\rm e}^{\rm hom}) \approx -0.5$ and continues with a steeper slope for high $\log(\Gamma_{\rm e}^{\rm hom})$, in good agreement with the observations. As our $\Gamma_{\rm e}$-dependent mass-loss prescription (Eq.\,\ref{eq:BEST}) was initially derived from the same data under the assumption that $\Gamma_{\rm e} = \Gamma_{\rm e}^{\rm hom}$ \citep[cf.\ ][]{bes1:14} this can be seen as a confirmation that most WNh stars in our sample are in fact evolving chemically homogeneously. 

Again, the overall agreement between models and observations in the WNh stage (with $\log(\dot{M}/\msunpyr) \gtrsim -5.2$) is very good, except for a short phase directly after the transition. In this short phase the surface is still more hydrogen-enriched than the stellar core, and thus $\Gamma_{\rm e} > \Gamma_{\rm e}^{\rm hom}$. However, as soon as the enhanced mass loss in the WNh phase sets in, the H-rich surface layers are quickly removed, and the stars become nearly chemically homogeneous.

While the O-star mass-loss rates in Fig.\,\ref{fig:GAMMA_VMS} (with $\log(\dot{M}/\msunpyr) \lesssim -5.2$) show a good agreement with observations for $\Gamma_{\rm e}^{\rm hom} \gtrsim -1$, the agreement gets worse for lower values of $\Gamma_{\rm e}^{\rm hom}$. For the O dwarfs in this regime \citet{sab1:17} derived upper limits for $\dot{M}$ which are indicated in grey. For O (super)giants \citet{ram1:17} provided explicit values for $\dot{M}$ that tend to be higher than those in our models. Notably, \citeauthor{ram1:17} did not provide absolute values for $D_{\rm mom}$ for the same objects (cf.\ Figs.\,\ref{fig:DMOM_VMS} and \ref{fig:DMOM_DUTCH}) because they considered the results as uncertain (H.\,Sana \& A.\,de\,Koter private communication). Depending on whether the mass-loss rates in this parameter range can be believed or not the differences between O dwarfs and (super)giants, which were already apparent in Fig.\,\ref{fig:DMOM_VMS}, may be even more pronounced for these objects. This is further supported by recent results of \citet{bes1:20,bes2:20} who investigated the mass-loss properties of the most massive stars in the R\,136 cluster in the centre of 30\,Dor. These authors determined a very young age of 1--2\,Myr for R\,136 and derived an $\dot{M}(\Gamma_{\rm e})$ relation which lies up to two orders of magnitude below the values derived for O stars the VFTS sample as indicated by the dashed curve in Fig.\,\ref{fig:GAMMA_VMS}.

\subsection{Location in the Hertzsprung-Russell Diagram}
\label{sec:HRD}

\begin{figure}[tbp!]
    \centering
    \includegraphics[scale=0.45,trim = 1cm 0. 0. 0.]{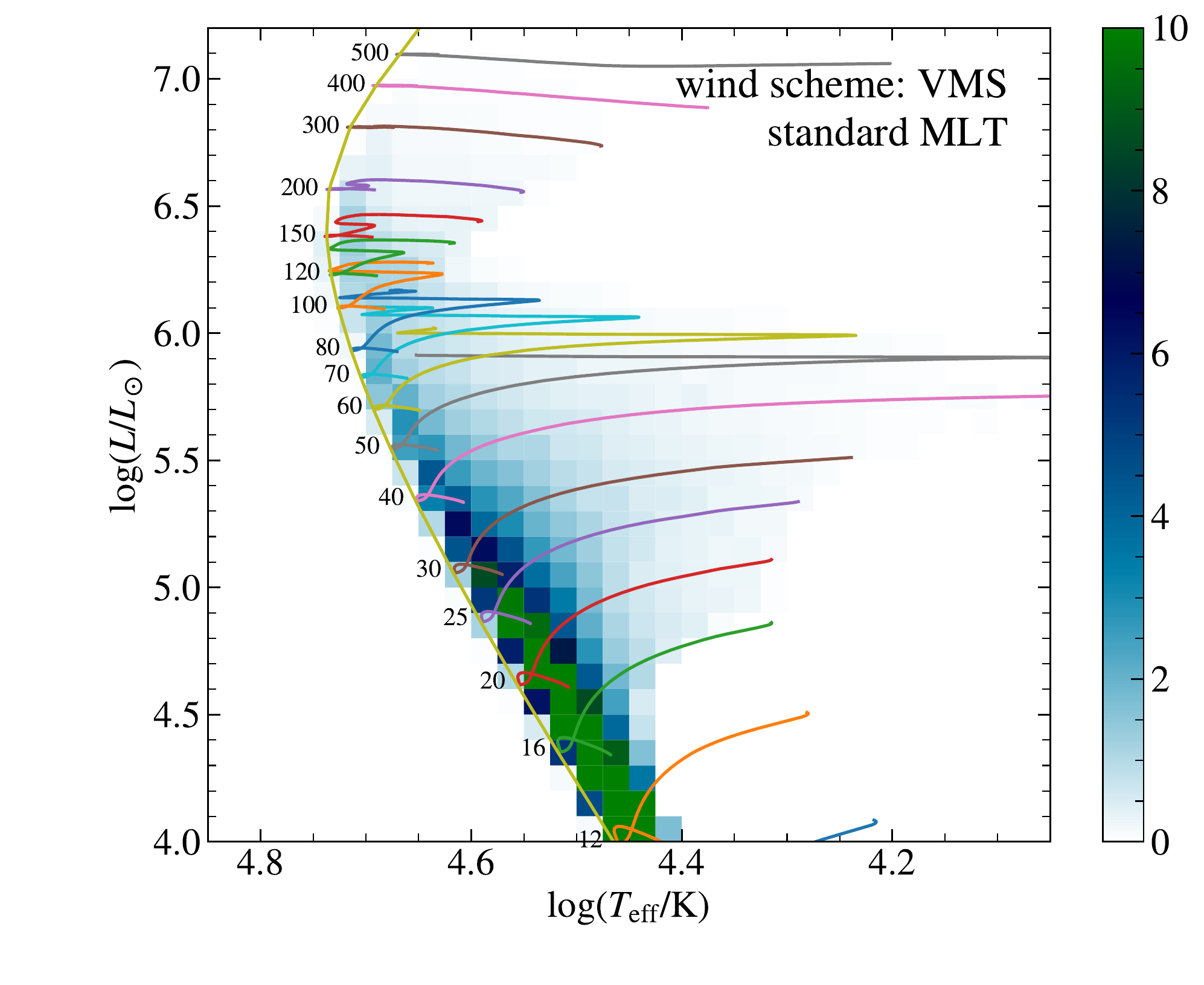}
    \caption{Theoretical Hertzsprung-Russell diagram for {\changed VMS models with standard MLT}. Coloured areas indicate the expected number of stars according to our population synthesis models for 30\,Dor (cf.\ Sect.\,\ref{sec:30DOR}). Evolutionary tracks for $\Omega/\Omega_{\rm crit} = 0.3$ are indicated, labelled with ZAMS masses in $M_\odot$.}
    \label{fig:HRD}
\end{figure}

\begin{figure}[tbp!]
    \centering
    \includegraphics[scale=0.45,trim = 1cm 0. 0. 0.]{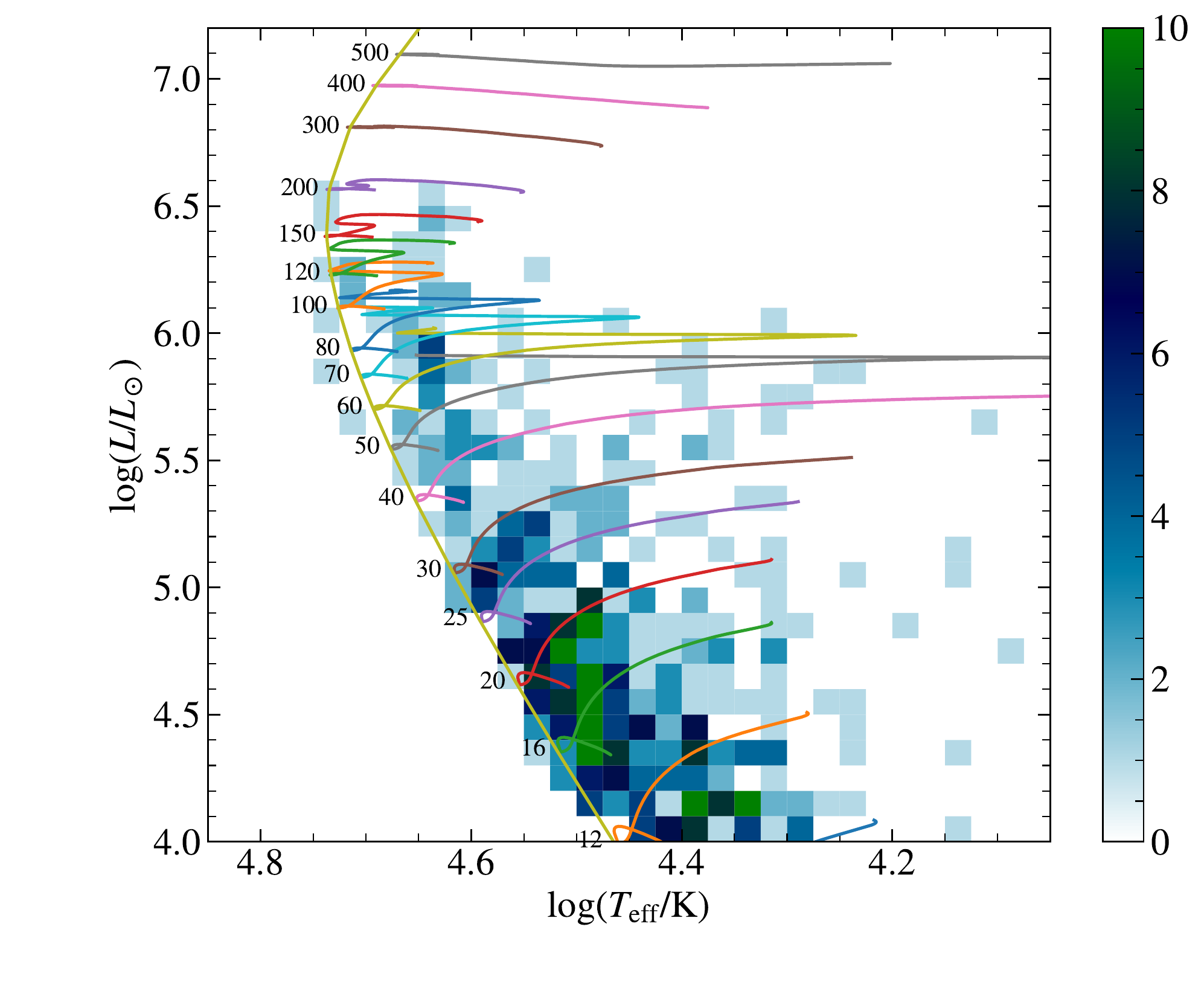}
    \caption{Observed Hertzsprung-Russell diagram for 30\,Dor. Coloured areas indicate the observed number of stars, evolutionary tracks are the same as in Fig.\,\ref{fig:HRD}.}
    \label{fig:OBS}
\end{figure}

\begin{figure}[tbp!]
    \centering
    \includegraphics[scale=0.45,trim = 1cm 0. 0. 0.]{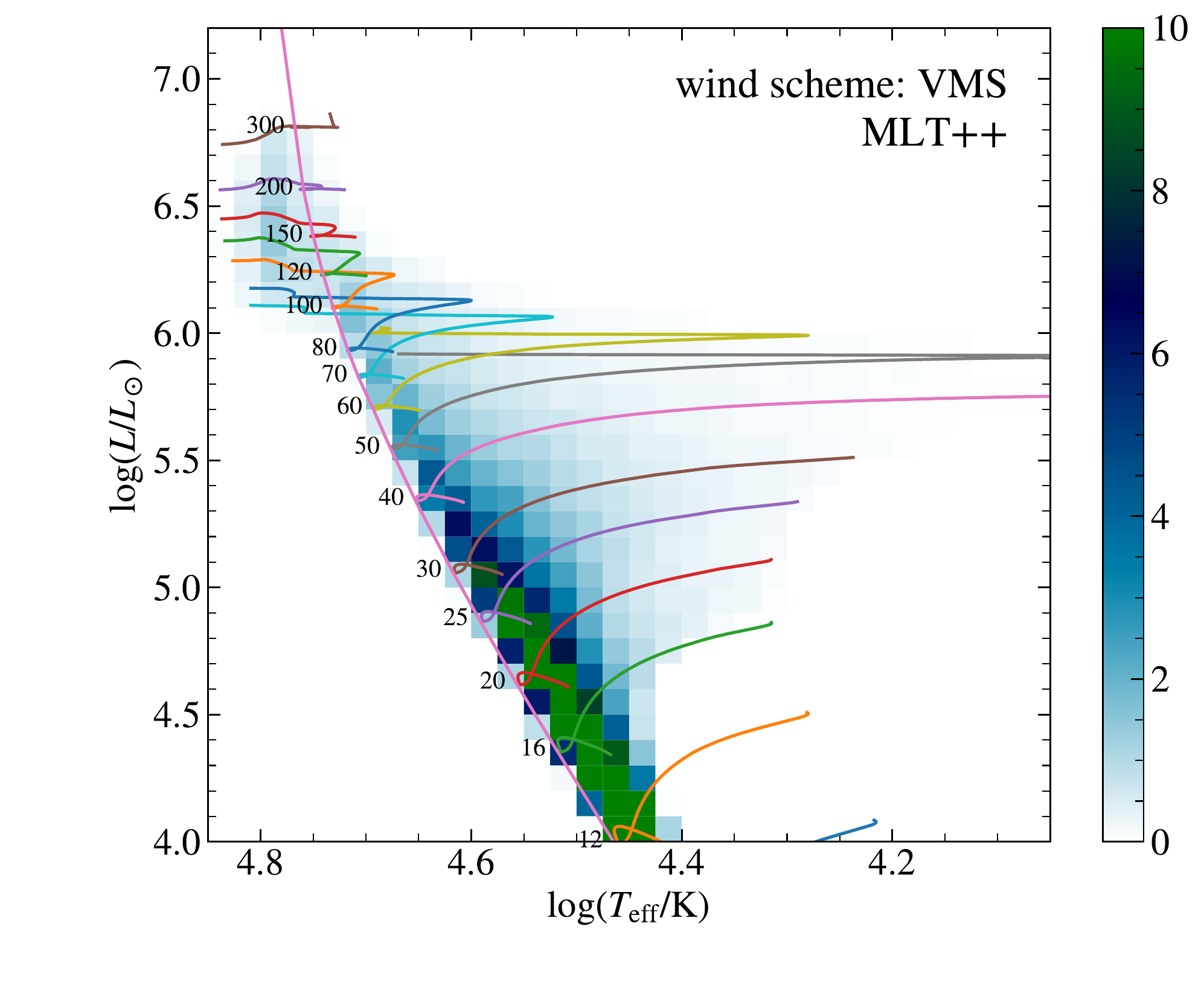}
    \caption{{\changed Influence of envelope inflation. Theoretical Hertzsprung-Russell diagram for models with reduced envelope inflation effect through an} artificially enhanced convective efficiency (MLT++; MESA option {\tt okay\_to\_reduce\_gradT\_excess = .true.}). }
    \label{fig:MLT++}
\end{figure}

In Fig.\,\ref{fig:HRD} we show the theoretical HRD for 30\,Dor including a colour coded map of the expected numbers of stars according to our population synthesis models. For comparison, we show the observed HRD in Fig.\,\ref{fig:OBS} with the observed number of stars colour coded in the same way. Generally, the observed HRD positions of stars above $\log(L/L_\odot) \approx 5$ are well reproduced by our VMS models. The lack of evolved stars below this luminosity in Fig.\,\ref{fig:HRD} is caused by the maximum age of 12\,Myr adopted in our population synthesis models. Beyond this, our VMS models reproduce the observed bending of the zero-age main sequence (ZAMS) for the highest masses, and the lack of stars with cooler temperatures than $\log(T_{\rm eff}/{\rm K})\approx 4.5$ above $\log(L/L_\odot) \approx 6$ very well.

The bending of the ZAMS towards cooler temperatures at the top of the main sequence is a consequence of the envelope inflation effect. This is demonstrated in Fig.\,\ref{fig:MLT++} where the inflation effect has been suppressed by artificially increasing the convective efficiency in radiation-dominated stellar envelopes (MESA option {\tt okay\_to\_reduce\_gradT\_excess = .true.}). As a consequence, the most massive stars above $\approx 80\,M_\odot$ become much hotter, and evolve towards even hotter temperatures during their main-sequence evolution. The reason is that stars in this regime become chemically homogeneous and thus more and more compact throughout their evolution. Only when standard MLT is used (Fig.\,\ref{fig:HRD}) the inflation effect compensates this effect and brings the effective temperatures of stars in this mass range in agreement with observations. 

\begin{figure}[t!]
    \centering
    \includegraphics[scale=0.45,trim = 1cm 0. 0. 0.]{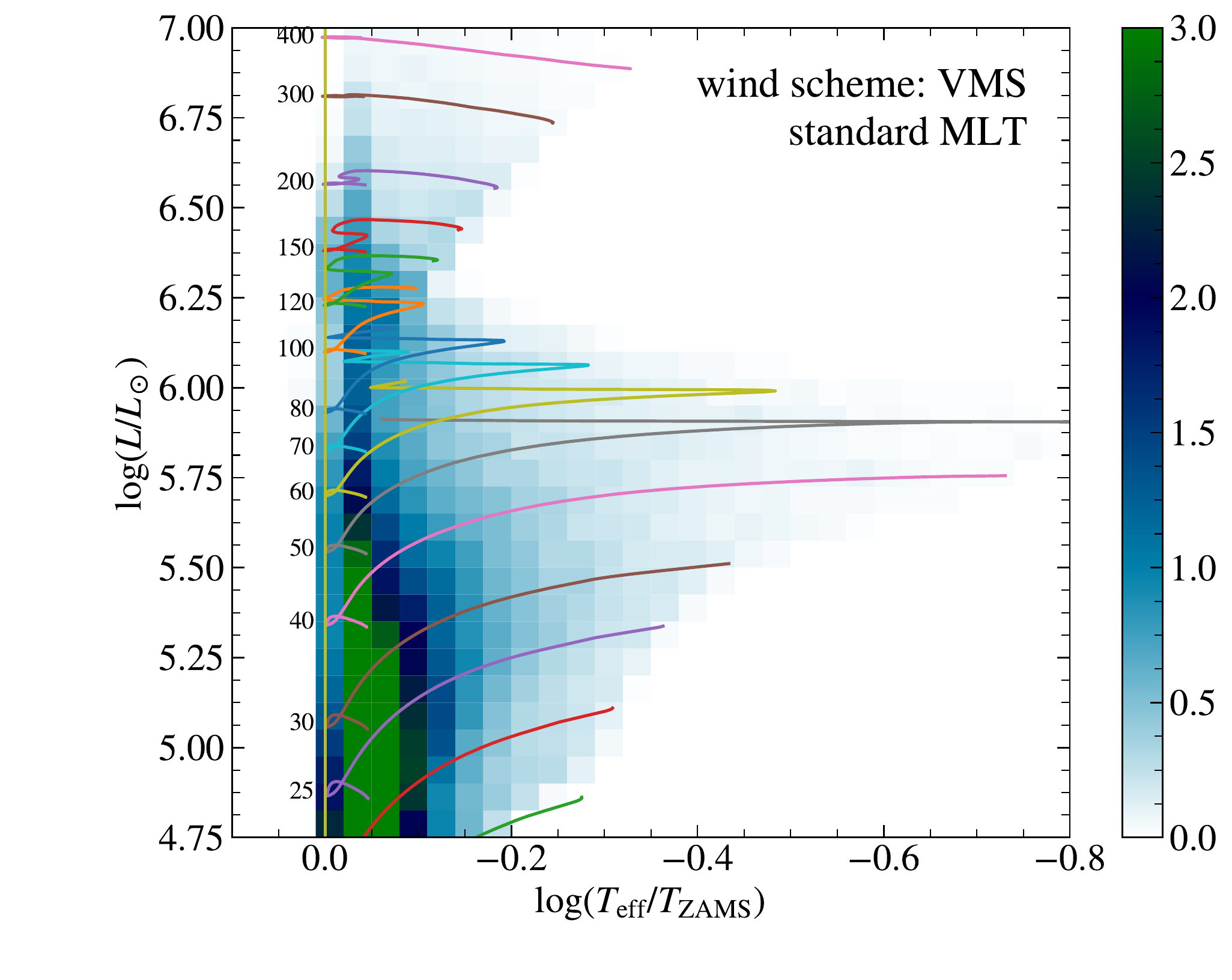}
    \caption{{\changed Effective temperature distribution.} Theoretical Hertzsprung-Russell diagram for the upper main sequence analogous to Fig.\,\ref{fig:HRD} but as a function of the effective temperature relative to the ZAMS temperature.}
    \label{fig:UHRD}
\end{figure}

\begin{figure}[t!]
    \centering
    \includegraphics[scale=0.45,trim = 1cm 0. 0. 0.]{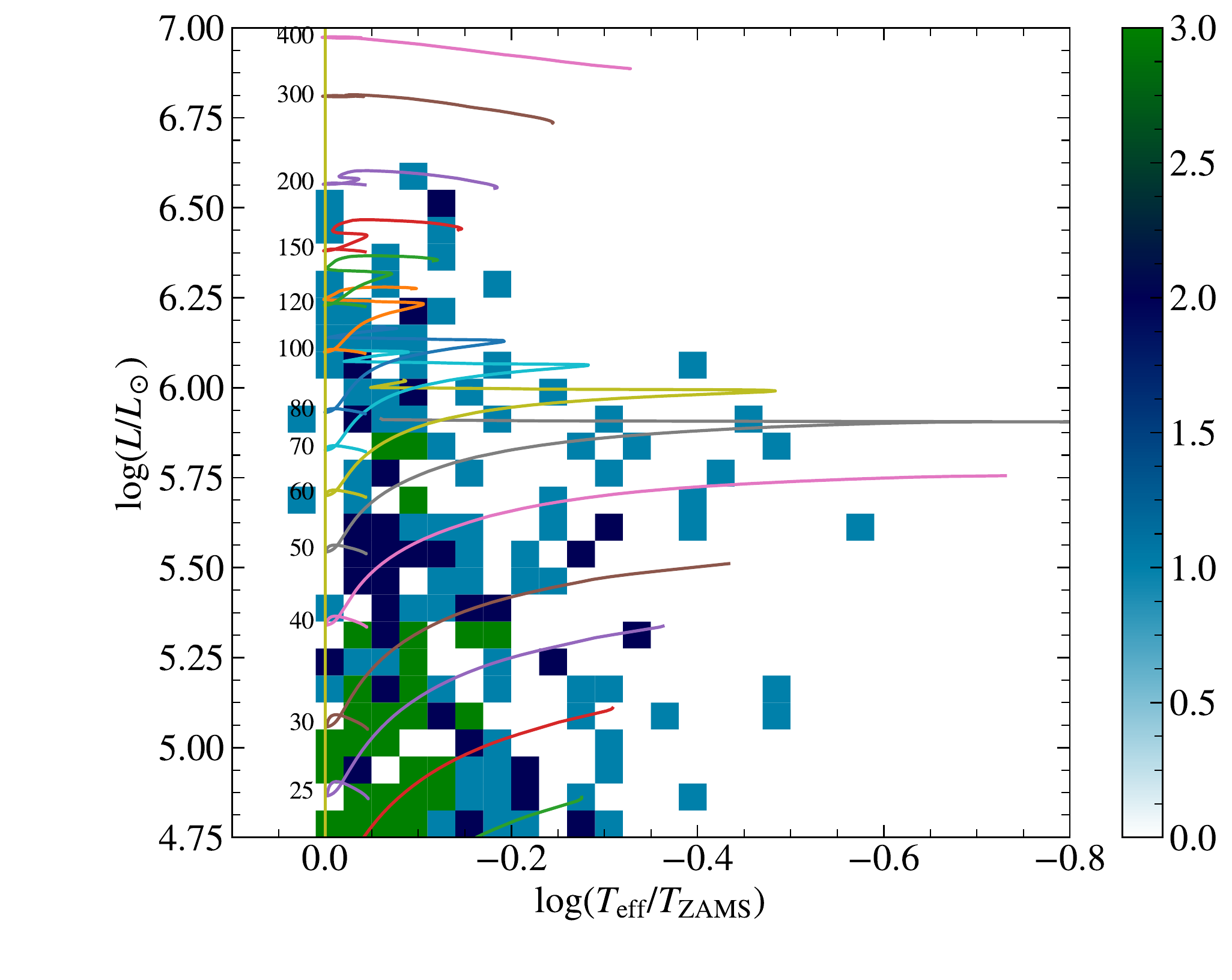}
    \caption{Observed Hertzsprung-Russell diagram analogous to Fig.\,\ref{fig:UHRD}.}
    \label{fig:UOBS}
\end{figure}

\begin{figure}
    \centering
    \includegraphics[scale=0.45,trim = 1cm 0. 0. 0.]{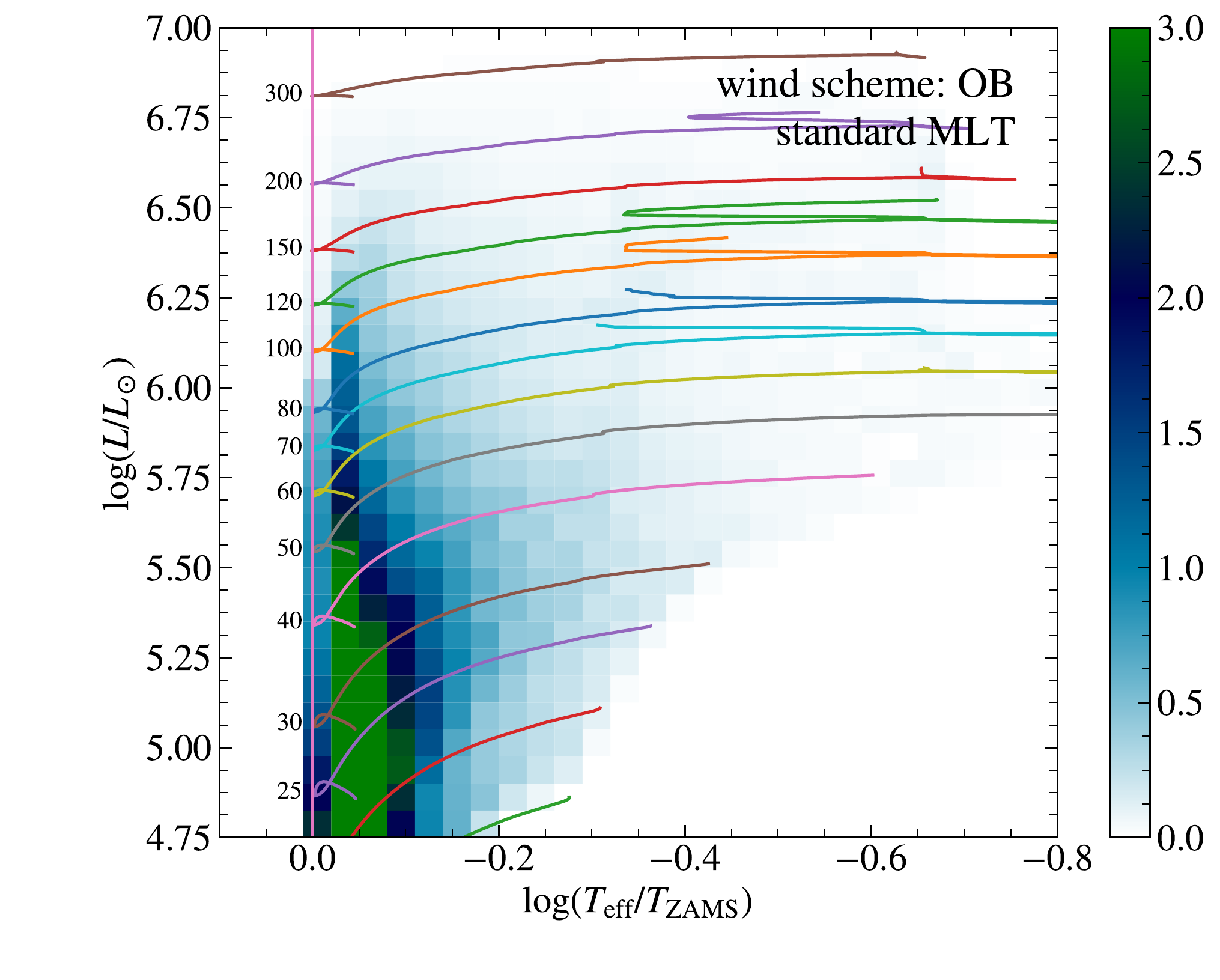}
    \caption{{\changed Influence of mass loss.} Theoretical Hertzsprung-Russell diagram analogous to Fig.\,\ref{fig:UHRD} but adopting a pure OB-type mass-loss prescription without enhanced $\Gamma$-dependent mass loss.}
    \label{fig:UHRD_OB}
\end{figure}

\begin{figure}
    \centering
    \includegraphics[scale=0.45,trim = 1cm 0. 0. 0.]{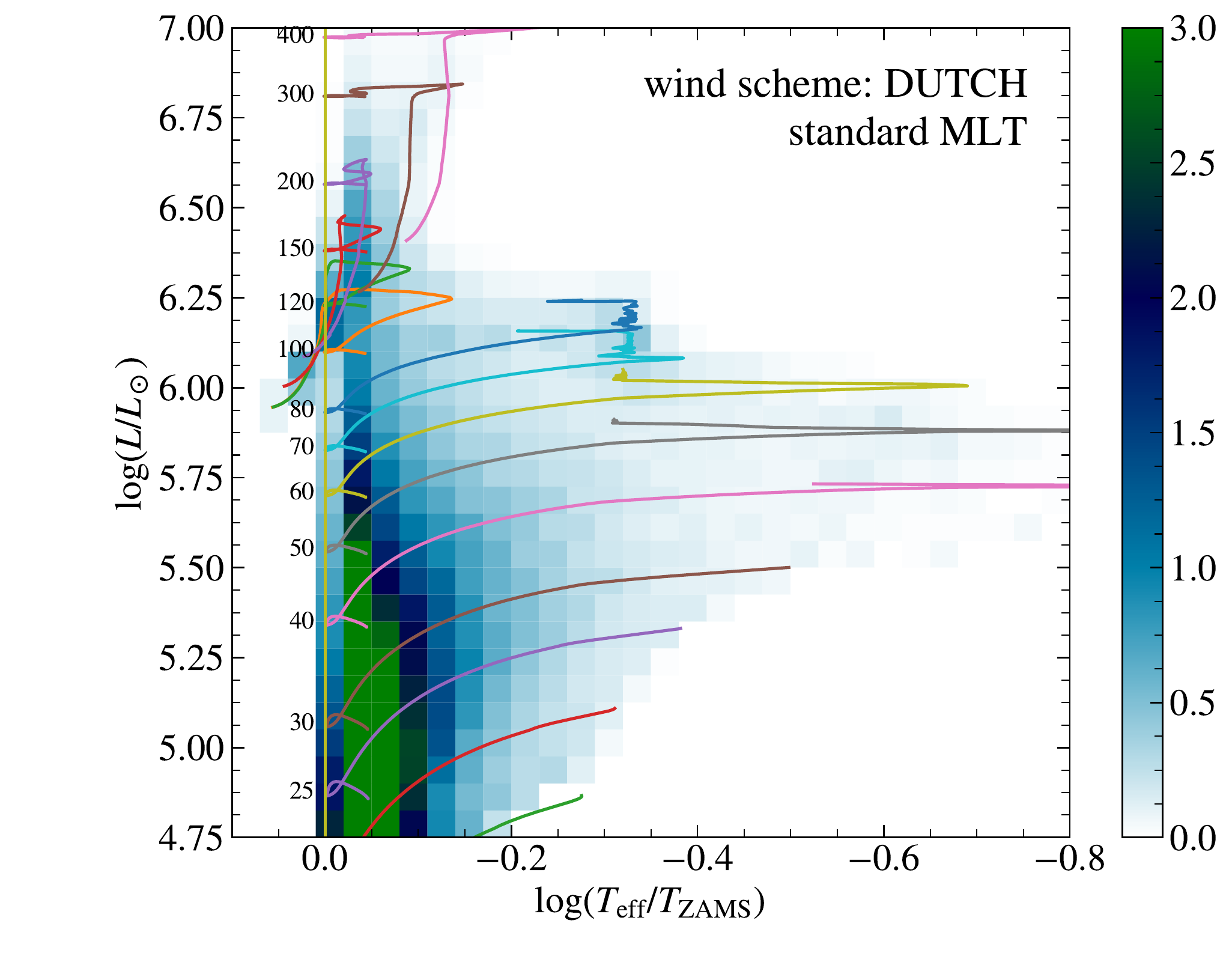}
    \caption{{\changed Influence of mass loss.} Theoretical Hertzsprung-Russell diagram analogous to Fig.\,\ref{fig:UHRD} but adopting the DUTCH MESA wind scheme.}
    \label{fig:UHRD_DUTCH}
\end{figure}

In Figs.\,\ref{fig:UHRD} and \ref{fig:UOBS} we focus on the distribution of stars in the upper HRD, relative to the ZAMS. To demonstrate how the predicted HRD positions depend on mass loss we show models with alternative mass-loss prescriptions in Figs.\,\ref{fig:UHRD_OB} and \ref{fig:UHRD_DUTCH}. In Fig.\,\ref{fig:UHRD_OB} the onset of the enhanced mass loss of WNh stars has been completely omitted while otherwise the same wind scheme as in our VMS models was used. Consequently, the mass loss of the most massive stars is reduced, and the outer H-rich layers are removed much more slowly, leading to higher hydrogen surface mass fractions $X_{\rm s}$ and therefore higher values of $\Gamma_{\rm e}$ (cf.\ Eq.\,\ref{eq:GAMMA_E}). Due to the strong dependence of the inflation effect on $\Gamma_{\rm e}$ \citep[cf.][Eqs.\ 20, 28, 29]{gra1:12}, this leads to an enhanced envelope inflation and correspondingly cooler stellar temperatures.

The models in Fig.\,\ref{fig:UHRD_DUTCH} are computed using the standard DUTCH wind scheme with much higher mass-loss rates than our VMS models. Because of the higher mass loss the most massive stars become more compact and less massive, and therefore hotter and less luminous. The qualitatively correct reproduction of the observed HRD positions of the most massive stars with $\log(L/L_\odot) \gtrsim 6$ by our VMS models can therefore be seen as an indirect confirmation of our new mass-loss prescription, and in particular of the adopted clumping factor $D$ in Eq.\,\ref{eq:BEST}.

For stars with $\log(L/L_\odot) \lesssim 6$ our models predict very extended envelopes and correspondingly cool stellar temperatures towards the end of the main-sequence evolution, irrespective of the mixing approach (cf.\ Figs.\,\ref{fig:HRD} and \ref{fig:MLT++}). In the observed HRD (Fig.\,\ref{fig:UOBS}) this region is populated by B-supergiants whose evolutionary origin is still unclear \citep[e.g.,][]{vin1:10,mce1:15}. In Fig.\,\ref{fig:UOBS} there are only few objects of this type that are clearly located in the Hertzsprung-gap, that is, outside the expected parameter range for main-sequence stars.  

\begin{figure}
    \centering
    \includegraphics[scale=0.53]{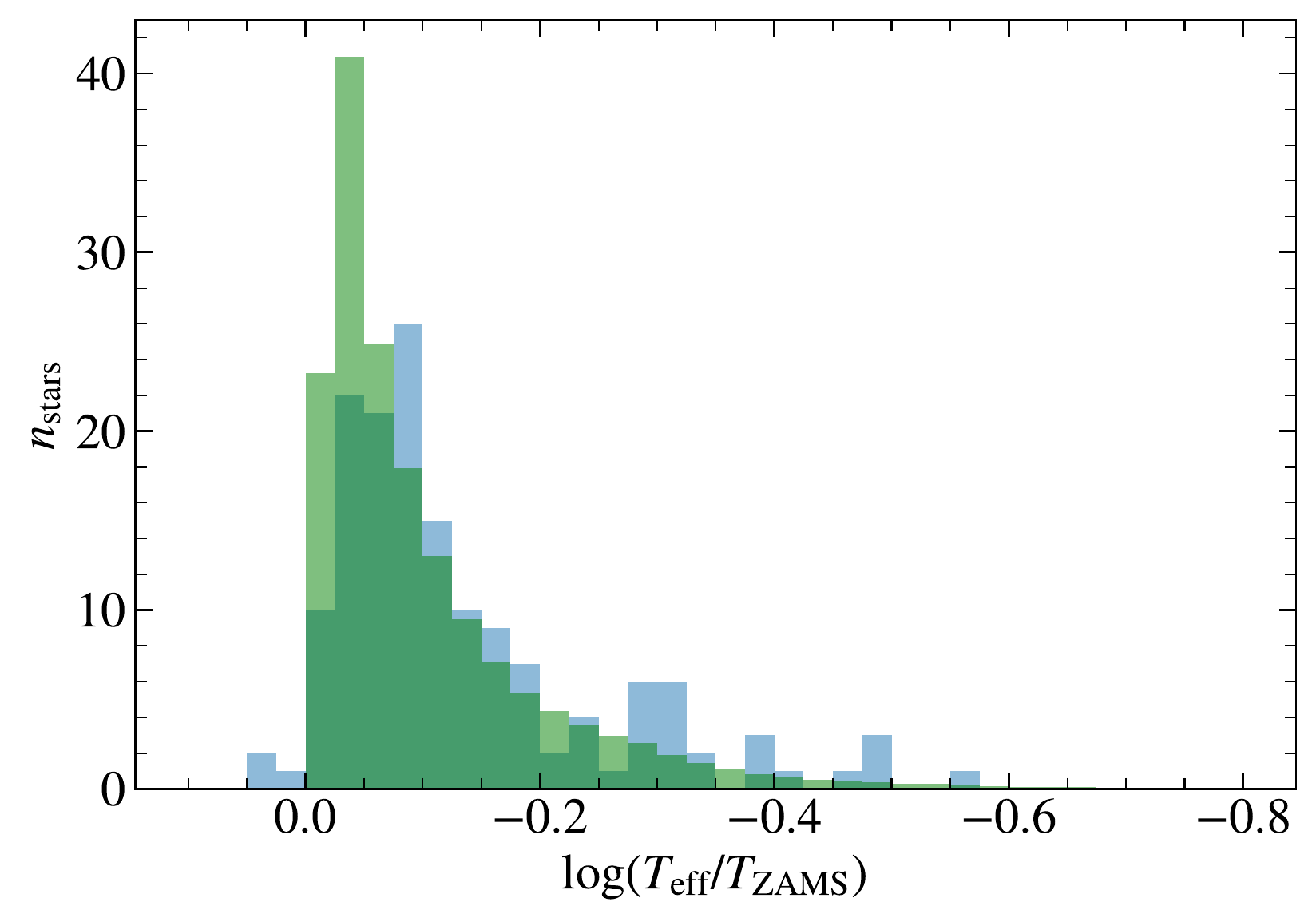}
    \caption{{\changed Effective temperature distribution.} Histogram of the observed numbers of single stars in the VFTS sample with $5 < \log(L/L_\odot) < 6$ as a function of their distance from the ZAMS ($\log(T_{\rm eff}/T_{\rm ZAMS})$) (transparent blue) compared with the theoretical distribution from our VMS population synthesis models (transparent green).}
    \label{fig:NvsT}
\end{figure}

To quantify the contribution of the envelope inflation effect we compare in Fig.\,\ref{fig:NvsT} the distribution of stars between $5 < \log(L/L_\odot) < 6$ as a function of their distance from the ZAMS ($\log(T_{\rm eff}/T_{\rm ZAMS})$) with observations. Our models predict 22.4 stars with $\log(T_{\rm eff}/T_{\rm ZAMS}) < -0.2$ compared to 30 stars that are observed. This means that we expect the majority of stars in this regime to be evolved main-sequence stars with inflated envelopes. Setting the maximum effective temperature for stars to be classified as B-type stars to 29\,kK, we predict 22.4 stars in this phase to appear as B supergiants, while 34 stars are observed. Again, according to our models the majority of B supergiants between $5 < \log(L/L_\odot) < 6$ are evolved main-sequence stars.

While the overall agreement between theory and observation in Fig.\,\ref{fig:NvsT} appears satisfactory, it is notable that there is a small systematic shift in $T_{\rm eff}/T_{\rm ZAMS}$ between the observed and predicted median locations of the main sequence, by 0.025\,dex. In principle, this could be an indication that the predicted core masses in our models are too small, similar to what has been claimed by \citet{tka1:20}. For this reason, we varied the overshooting parameter analogous to \citeauthor{tka1:20} as well as the parameters for rotational mixing in our models but did not achieve a satisfactory agreement between theory and observations. Also, the relatively high oxygen abundance in our models \citep[based on the solar composition from][]{gre1:98} does already bring our models towards slightly lower effective temperatures than what would result from more recent estimates \citep[such as][]{asp1:09}. Therefore, it remains unclear whether the observed shift between theory and observations is due to deficiencies in our models or systematic errors in the spectroscopic analyses of the VFTS sample.

\subsection{Helium surface enrichment}

\label{sec:HE}

\begin{figure}[t]
    \centering \includegraphics[scale=0.45]{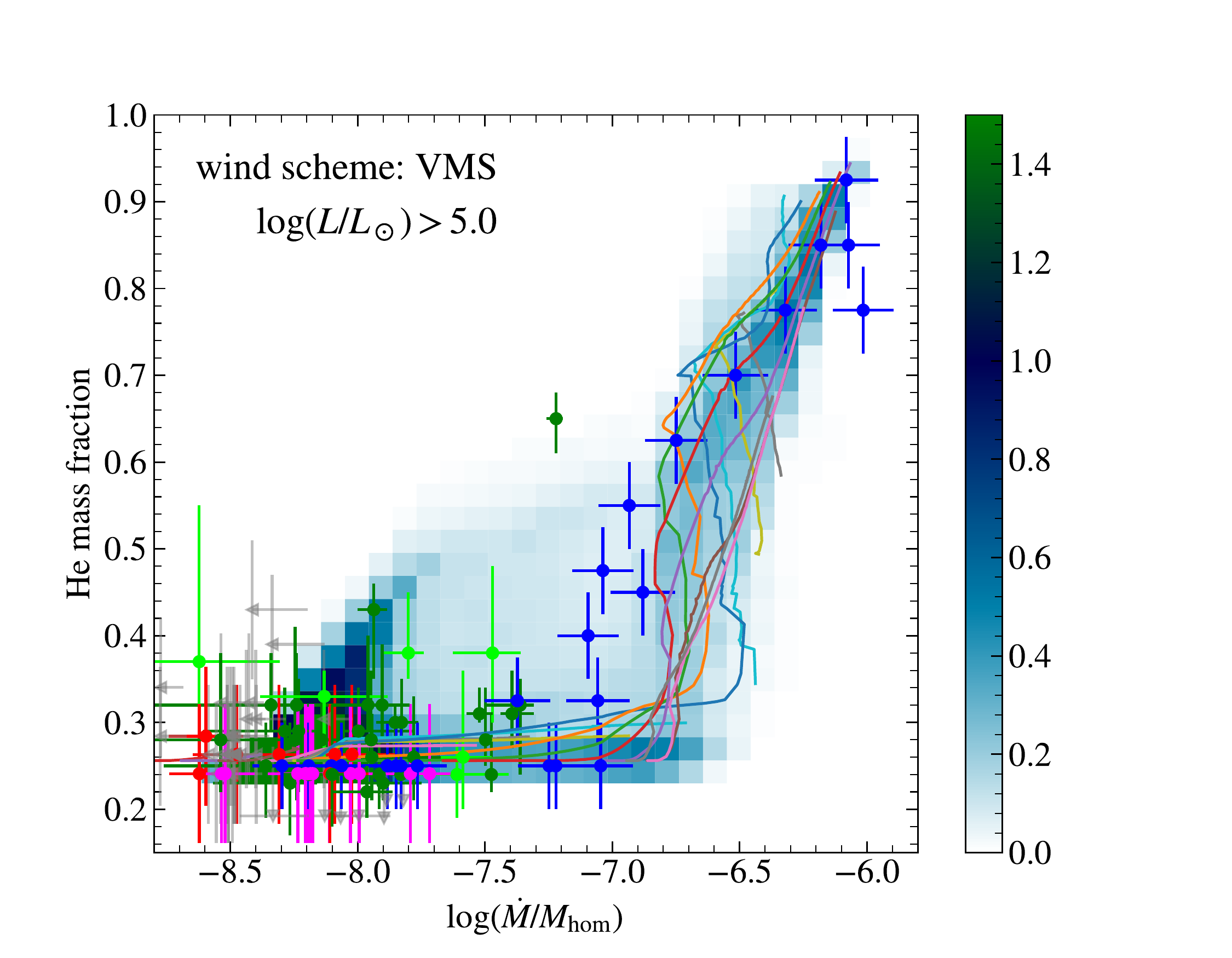}
    \caption{{\changed Helium surface enrichment through mass loss.} Helium surface mass fraction $Y$ as a function of the estimated inverse mass-loss timescale $\dot{M}/M_{\rm hom}$ for stars with $T_{\rm eff} > 30$\,kK and $\log(L/L_\odot)>5$, compared with VFTS O and WNh stars labelled as in Fig.\,\ref{fig:DMOM_VMS}. The colour scale indicates the number of objects predicted by our population synthesis. Evolutionary tracks from Fig.\,\ref{fig:HRD} are indicated by coloured lines.}
    \label{fig:XHE}
\end{figure}

In Fig.\,\ref{fig:XHE} we show a comparison between the observed and predicted
helium surface mass fractions $Y_{\rm s}$ for stars with $T_{\rm eff} >
29$\,kK, i.e., for O-type stars and WNh stars in the VFTS, as a function of
their estimated inverse mass-loss timescale $\dot{M}/M_{\rm hom}$. Again, for
a meaningful comparison between models and observations, the mass estimate
$M_{\rm hom}(L,X)$ from \citet{gra1:11} has been computed in an identical manner for models and observations.  

Our population synthesis models show two regions with substantially enhanced values of $Y_{\rm s}$ in Fig.\,\ref{fig:XHE}. The WNh stars with $\log{\dot{M}/M_{\rm hom}} \gtrsim -7.5$ follow a clear trend of increasing $Y_{\rm s}$ for increasing $\dot{M}/M_{\rm hom}$ \citep[cf.\ also][]{bes1:14,bes2:20}. The reason is that the enhanced mass loss in the WNh regime efficiently removes the hydrogen rich outer layers, leading to quasi chemically homogeneous evolution with a strong dependence of $\dot{M}$ on $\Gamma_{\rm e}$, and therefore on $Y_{\rm s}$.

\begin{figure}[t]
    \centering
    \includegraphics[scale=0.45]{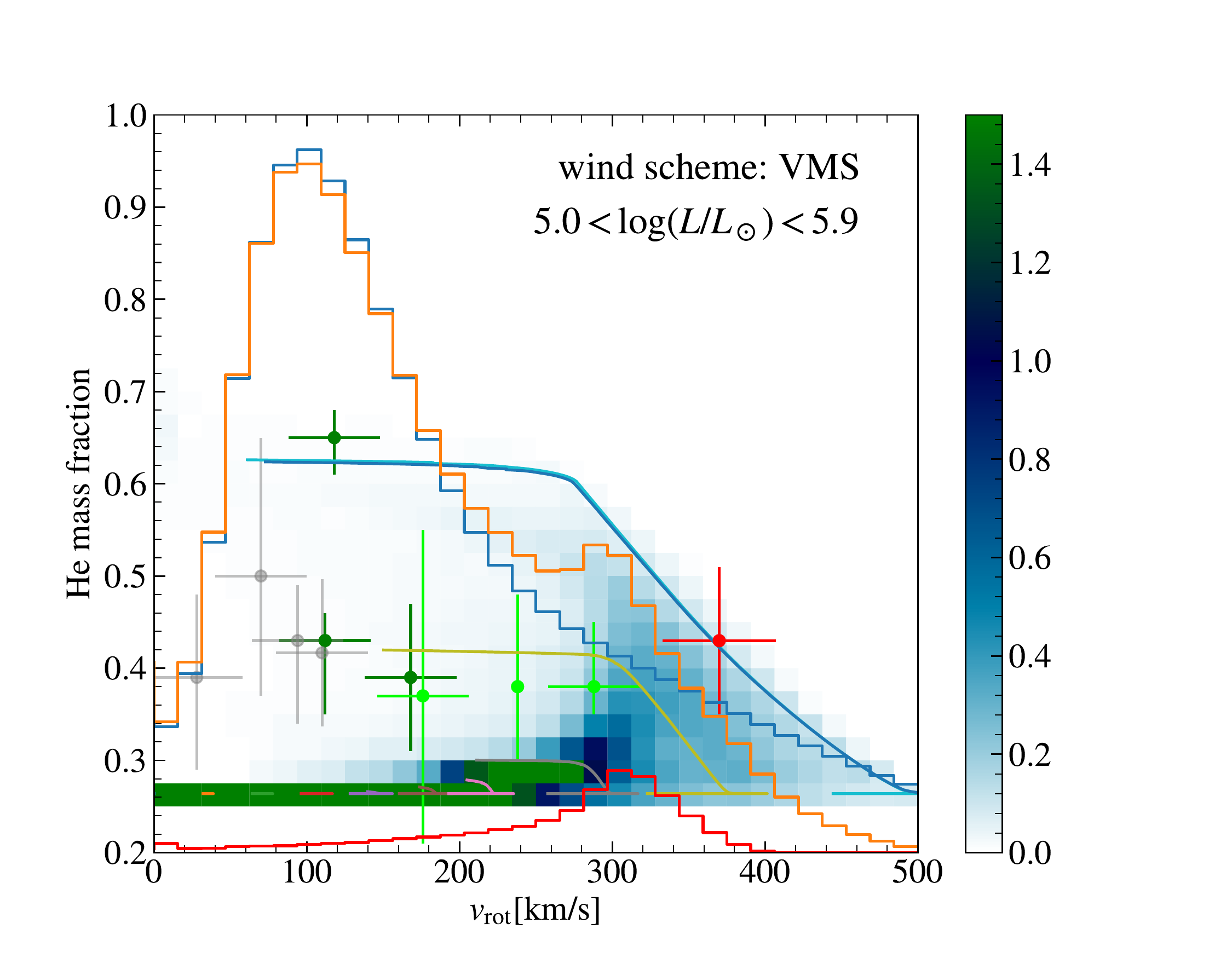}
    \caption{{\changed Helium surface enrichment through rotation.} Helium surface mass fraction $Y_{\rm s}$ as a function of rotational velocity for {\changed O}\,stars with $T_{\rm eff} > 29$\,kK and $5.0 < \log(L/L_\odot) < 5.9$, compared with observed $Y_{\rm s}$ and $\varv \sin i$  of He-enriched VFTS O stars (with $Y_{\rm s}>0.35$) labelled as in Fig.\,\ref{fig:DMOM_VMS}; stars with $\log(L/L_\odot)<5.0$ are plotted in grey. The colour scale indicates the number of objects predicted by our population synthesis models. Evolutionary tracks for an initial mass of 30\,$M_\odot$ are indicated by coloured lines. Step functions indicate the initial (blue) and present (orange) total number distributions in arbitrary units, as well as the present distribution of He-enriched stars with $Y_{\rm s}>0.35$ (red).}
    \label{fig:ROT}
\end{figure}

In the O-star range, with lower values of $\dot{M}/M_{\rm hom}$, our models
predict substantial He enrichment for initially fast rotating stars. This is
demonstrated in Fig.\,\ref{fig:ROT} where we show the He enrichment for O
stars with $5 < \log(L/L_\odot) < 5.9$ (i.e., excluding WNh stars) as a
function of rotational velocity. According to our models, fast rotating stars
in this regime develop strong winds {\changed due to their proximity to the
$\Gamma\Omega$-limit}. This leads to a phase of enhanced mass loss which removes
angular momentum until the rotational velocity falls below a value of
typically $300$\,km/s. As a consequence, there is a peak in the rotational
velocity distribution for O stars at $\varv_{\rm rot} \approx 300$\,km/s
(cf.\ the orange step function in Fig.\,\ref{fig:ROT}). Notably, \citet{ram1:13} found a similar feature in the observed rotational velocity distribution of the VFTS O\,star sample, however, at a higher velocity of $\approx 400$\,km/s.

According to our models the rotational braking associated with this process
induces substantial mixing of helium-rich material to the surface, with a
maximum in the number distribution of He-enriched stars at $\approx 300$\,km/s
(cf.\ the red step function in Fig.\,\ref{fig:ROT}). The same objects
constitute the group of helium enriched stars with $\log{\dot{M}/M_{\rm hom}}
\lesssim -7.5$ in Fig.\,\ref{fig:XHE}. According to our models we expect 14.6 O\,stars with $Y_{\rm s} > 0.35$ in the luminosity range $5.0 < \log(L/L_\odot) < 5.9$ in the VFTS sample which have been enriched by rotational mixing. This compares to 7 observed objects in the same regime which are indicated by coloured symbols in Fig.\,\ref{fig:ROT}. For lower luminosities we expect 0.7 objects, compared to 5 observed objects which are indicated by grey symbols in Fig.\,\ref{fig:ROT}. 

Notably, VFTS\,285, the star with the highest value of $\varv \sin i = 600$\,km/s in our sample and $Y_{\rm s} = 0.36$ belongs to the group of low-luminosity stars for which no He-enhancement is predicted (VFTS\,285 is not visible in Fig.\,\ref{fig:ROT}). In both, the high- and low-luminosity regime, we expect the vast majority of He-enriched objects near a rotational velocity of $\approx 300$\,km/s. Due to the effect of random inclination angles the expected average value of $\varv \sin i$ is $\pi/4$ \citep[cf.][]{bro1:11}, with a decreasing distribution function for small values of $\sin i$. We therefore expect the distribution of He-enriched stars to peak near 235\,km/s, while the observed $\varv \sin i$ distribution appears evenly distributed, with a potential peak near $100$\,km/s. This means that neither the predicted numbers of rotationally enriched stars, nor their distribution agrees with observations. The implementation of rotational mixing and/or rotationally induced mass loss in our models therefore seems to leave space for improvements (cf.\ the discussion in Sect.\,\ref{sec:HE_disc}).

\section{Discussion}
\label{sec:DISC}

In the following we discuss the implications of our results for the physics and evolution of the most massive stars (Sect.\,\ref{sec:WNh_disc}), their upper mass limit (Sect.\,\ref{sec:ulim}), the evolution of OB stars (Sect.\,\ref{sec:OB_disc}), and the mechanisms for surface He-enrichment in the upper HRD (Sect.\,\ref{sec:HE_disc}).

\subsection{WNh stars: the most massive stars}

\label{sec:WNh_disc}

According to our analysis, at LMC metallicity, stars in excess of $\approx 80\,M_\odot$ enter the WNh stage at some point during their main-sequence evolution. This means that they reach high enough Eddington factors to initiate optically thick winds, so that they appear as WR-type emission-line stars. Because they still have hydrogen on their surface, they are traditionally classified as H-rich WN stars, or WNh stars. As shown in Sect.\,\ref{sec:MDOT30}, the absolute mass-loss rates, as well as the onset of WR-type mass-loss in this regime, are well described by our VMS models employing a $\Gamma_{\rm e}$-dependent mass-loss relation.

Because of their enhanced mass loss, the outer radiative layers of WNh stars are quickly removed so that they develop a quasi chemically homogeneous structure, with hydrogen surface abundances starting at near-solar values and almost reaching hydrogen depletion at the end of their main-sequence evolution. This is in stark contrast to the traditional treatment of WR-type mass loss in stellar evolution models, where enhanced mass loss only sets in beyond a limit of $Y_{\rm s} > 0.4$.  In Sect.\,\ref{sec:HE} we have shown that our VMS models describe the resulting He-enrichment through the enhanced mass loss in the WNh stage well, in good agreement with observations.

Because of their proximity to the Eddington limit, WNh stars are subject to the envelope inflation effect. Therefore, their HRD positions are shifted towards cooler temperatures. For masses $\gtrsim 80\,M_\odot$ the ZAMS is bent towards cooler temperatures due to this effect. As we have shown in Sect.\,\ref{sec:HRD}, our VMS models with standard MLT reproduce this effect very well, in agreement with the observed HRD positions. In particular, there appears to be no need to invoke a clumped envelope structure to explain their radii as it has been proposed for the inflated envelopes of classical WR stars \citep{gra1:12,gra2:13}.

A further consequence of the inflation effect is that WNh stars evolve towards cooler temperatures, i.e., towards the right side in the HRD, although they are close to chemical homogeneity. Again, the HRD positions following from our VMS models agree with observations. Models that employ the oft-used modified treatment of super-adiabatic convection in radiation-dominated regions \citep[MLT++,][]{pax1:13} clearly fail to reproduce the observed HRD positions of WNh stars. They show much hotter temperatures than observed and evolve towards the left side in the HRD (cf.\,Figs.\ \ref{fig:OBS} and \ref{fig:MLT++}). 

The general agreement of our models with the observed population of WNh stars in 30\,Dor in terms of numbers, HRD positions, mass loss and surface enrichment strongly support their evolutionary status as VMS in the phase of core H-burning.

\subsection{A variable upper stellar mass limit}

\label{sec:ulim}

The first estimate of an upper limit for the masses of stars of $\approx$\,$150\,M_\odot$ was determined by \citet{fig1:05}, based on photometric data of the Arches cluster in the Galactic Centre, one of the youngest and densest clusters in the Galaxy. Later \citet{cro1:10} analysed the brightest WNh stars in R\,136, the central cluster of 30\,Dor, and found luminosities up to $\log(L/L_\odot) = 6.94\pm0.1$, corresponding to initial masses up to $300\,M_\odot$. Because of its high stellar density R136 has not been included in the VFTS sample but \citet{bes2:20} analysed the 55 brightest stars in R\,136 based on new HST spectroscopy of \citet{cro1:16} and revised the upper luminosity limit to $\log(L/L_\odot) = 6.8\pm0.1$, corresponding to an upper mass limit of $\approx 250\,M_\odot$.

According to our present models the brightest WNh stars evolve with near-constant luminosity, implying slightly higher initial masses than the ones derived by \citeauthor{bes2:20}\ As a consequence, a value of $\log(L/L_\odot) = 6.8$ corresponds almost precisely to $M_{\rm ini} = 300\,M_\odot$. As we have shown in Sect.\,\ref{sec:NvsL}, the VFTS sample displays an upper luminosity cut-off at a slightly lower value of $\log(L/L_\odot)\approx 6.6$ (cf.\ Fig.\,\ref{fig:NvsL}), corresponding to an initial mass of $\approx 200\,M_\odot$. According to our population synthesis models we would expect $4.7$ stars above $\log(L/L_\odot)=6.6$ and 1.8 stars above $\log(L/L_\odot)=6.8$ without assuming an upper mass limit. If the R\,136 upper mass limit would be universal, we would therefore miss 3.1 stars with $6.6 < \log(L/L_\odot) < 6.8$ in the VFTS sample. Given that there are 20 stars with $\log(L/L_\odot)>6.0$, the probability to miss 3 stars with $6.6 < \log(L/L_\odot) < 6.8$ by chance would be $(17/20)^{20} = 0.039$. If there were no uncertainties in the individual luminosity measurements this would indicate a significant difference of the upper mass limits in R\,136 and the remainder of 30\,Dor.

Now the relevant most massive stars in both samples are spectroscopically remarkably similar objects of subtype WN5h which have been analysed by \citet{bes1:14} for the VFTS sample and \citet{bes2:20} for R\,136. In both works individual uncertainties of $\pm 0.1$\,dex in $\log(L)$ were estimated which are of course significant compared to the difference in the maximum luminosities of 0.2\,dex. However, we think that the {\em relative} uncertainties in the derived $\log(L)$ may be smaller than the individual errors because of the similarity of the objects, and because they have been analysed by the same person using the same method of analysis. Moreover, the brightest star in the VFTS sample, VFTS\,1025 alias R\,136c is in fact a member of R\,136 with a luminosity of $\log(L/L_\odot) = 6.58$. The second brightest star, VFTS\,682 has a much lower luminosity of $\log(L/L_\odot) = 6.51$, and has been discussed as a slow runaway from R\,136 \citep{ren1:19}. In any case the VFTS population excluding R\,136c clearly displays lower luminosities than $\log(L/L_\odot) = 6.6$ compared to R\,136 with its brightest member R\,136a1 with $\log(L/L_\odot) = 6.78$.

As \citet{bes2:20} report 3 stars with $6.6 < \log(L/L_\odot) < 6.8$ in R\,136, in agreement with expectations for a top-heavy IMF, dynamical mass segregation cannot account for the lack of the same number of stars in the VFTS sample. Instead, the higher mass limit for R\,136 could imply that R\,136 initially formed stars with higher masses, depending on the local conditions in the star-forming region. This could mean that the stars in 30\,Dor and R\,136 already formed in a mass-segregated way \citep[cf.][]{pav1:19}. Alternatively, the most massive stars in R\,136 could be the result of successive stellar mergers due to the high stellar density in the cluster \citep[e.g.][]{por1:04}.

\subsection{OB stars}
\label{sec:OB_disc}

\subsubsection{Mass-loss rates}

For the mass-loss rates of OB stars we employed a modified form of the \citeauthor{vin1:01} recipe adjusted to match the empirical mass-loss rates in the VFTS sample with an adopted clumping factor of $D=10$. To this purpose we adopted a shallower dependence on $L$ from \citet{bes1:14} and used overall reduced mass-loss rates. In Sect.\,\ref{sec:MDOT30} we have shown that this mass-loss relation does indeed match the average O\,star mass-loss rates in the VFTS sample, however, it fails to reproduce their observed wide spread. In particular, the O\,dwarfs analysed by \citet{sab1:17} show systematically lower wind momenta and mass-loss rates than the O\,(super)giants analysed by \citet{ram1:17}. This trend of reduced mass-loss rates for younger objects becomes even more prominent when the VFTS results are compared with the mass-loss relation from \citet{bes1:20} for O and WNh stars in R\,136. The stars in R\,136 are only about 2\,Myrs old, and therefore even younger than most of the stars in the VFTS sample. Notably, the mass-loss rates for O\,stars in R\,136 are up to two orders of magnitudes lower than those found in the VFTS sample (cf.\ Fig.\,\ref{fig:GAMMA_VMS}), possibly indicating a very strong trend of reduced mass-loss rates for young O dwarfs. 

A reduction of O-star mass-loss rates is also reported in recent theoretical works that employ more elaborate techniques to compute the radiative force than the oft-used first-order Sobolev approximation as in \citeauthor{vin1:01}\ Employing higher-order Sobolev models \citet{owo1:99} reported a drop in the line source function near the sonic point due to rapid changes in the velocity gradient. \citet{gra4:03} reported a similar effect in non-LTE models employing a co-moving frame (CMF) radiative transfer causing a local drop in the radiative force near the sonic point. More recent mass-loss predictions for OB stars employing higher-order Sobolev or CMF techniques by \citet{luc1:07,krt1:17,sun1:19} report similar effects and predict much lower mass-loss rates than those of \citeauthor{vin1:00}\ One could speculate that young stars with high $\log g$ are more susceptible to this effect because a high $\log g$ implies stronger changes in the velocity gradient. Physical effects occurring in more evolved stars that are not included in current wind models, such as turbulence caused by sub-surface convection \citep{can1:09}, could help to overcome the associated local drop in the radiative force for these objects, and help to bring their mass-loss rates closer to the values of \citeauthor{vin1:00} as it is observed for the O supergiants in the VFTS sample \citep[cf.][]{ram1:17}.

\subsubsection{HRD positions}

In agreement with previous studies by \citet{koe1:15, san1:15} we find that luminous OB stars are subject to the inflation effect, albeit less dependent on the detailed treatment of convection than for the more luminous WNh stars. As a consequence, such objects appear as B\,supergiants during their main-sequence evolution, in regions of the HRD that previously may have been attributed to the Hertzsprung-gap. In Sect.\,\ref{sec:HRD} we have shown that, according to our population synthesis models, 22 out of the 34 bright B\,supergiants in the VFTS sample are expected to be inflated main-sequence stars. However, there are some (at least 4) objects in the VFTS sample whose HRD positions cannot be explained by our models, so that alternative explanations such as blue loops or binary interaction may still be needed to explain the complete B\,supergiant population.

Notably, our population synthesis models are also able to reproduce the lack of B\,supergiants with high luminosities ($\log(L/L_\odot) \gtrsim 6$) in the VFTS sample. In Sect.\,\ref{sec:HRD} we identified the removal of H-rich layers due to the onset of enhanced WR-type mass loss as the reason for this effect.  The successful reproduction of the VFTS population in the HRD is therefore also an indirect confirmation of the absolute mass-loss rates and clumping factors adopted in our models.

Previous works such as \citet{gra1:12,gra1:18} associated stars with inflated envelopes with the S\,Doradus type variability of Luminous Blue Variables (LBVs). The fact that we expect about 20 objects in the VFTS sample to be in this phase, but none of the observed stars shows LBV-type behaviour, implies that envelope inflation is not {\em generally} associated with the LBV phenomenon, in particular not in the H-burning phase. This could still mean that the inflation effect is associated with LBVs, but under more rare conditions such as fast rotation, or in very late (pre-SN) evolutionary stages \citep[cf.][]{gro1:06,gra1:12,gro1:13}.

Finally, a clumped density structure inside inflated stellar envelopes has been proposed by \citet{gra1:12,gra2:13} to explain the observed radii of classical core He-burning WR stars. Our present analysis shows that there is no need to invoke similar clumped sub-surface regions to explain the radii of B supergiants through the inflation effect.

\subsection{Surface helium enrichment through mass loss and rotation}
\label{sec:HE_disc}

As outlined in Sect.\,\ref{sec:WNh_disc} the surface He-enrichment through the strong mass loss of WNh stars is well reproduced by our models. On the other hand, the He-enrichment through rotational mixing appears to be over-predicted by our models. According to our models we expect to find 15 O stars with $Y_{\rm s} > 0.35$ in the VFTS sample that are He-enriched through rotational mixing. These objects are expected to have high rotational velocities ($\approx 300$\,km/s) and high luminosities $5.0 < \log(L/L_\odot) < 5.9$. In reality, we find 12 He-enriched objects that are concentrated at low values of $\varv \sin i$ of $\approx 100$\,km/s, five of which are located at luminosities of $\log(L/L_\odot) < 5.0$. Therefore, the observed sample appears to be largely incompatible with our model predictions, implying a much lower number of truly rotationally enriched O stars.

This result implies that either the internal rotational mixing is over-predicted in our models, or that the rotationally enhanced mass loss, which induces mixing, is too high. A large part of the 15 He-enriched stars in the VFTS sample could therefore be false detections (such as un-identified multiple systems) or products of alternative processes such as binary interaction or mergers \citep[cf.][]{dem1:14}. Whether the observed discrepancy can be explained through changes in internal mixing parameters, or the previously discussed lower mass-loss rates for young O stars, remains to be shown in future works.

\section{Conclusions}
\label{sec:CON}

The implementation of new mass-loss relations in combination with standard-MLT envelope physics in our present evolutionary models leads to a successful reproduction of the properties of the most massive stars ($\gtrsim 80$--$100\,M_\odot$) in 30\,Dor, including their HRD positions, mass-loss rates, and surface He-enrichment. In particular the transition between optically thin O-star winds and the optically thick winds of WNh stars is well reproduced.  

For the less massive O\,stars (with $\approx25$--$100\,M_\odot$) we find much more diverse mass-loss properties than previously thought, with indications for substantially reduced mass-loss rates for young O dwarfs. Also, the surface helium enrichment in the O-star range does not appear to be well understood, as we find no clear indications for rotational mixing of helium in contrast to our model predictions. Both phenomena may be related as reduced mass-loss rates for young fast-rotating stars could lead to less rotational braking and mixing.

According to our models the distribution of stellar luminosities in 30\,Dor is in good agreement with the top-heavy IMF from \citet{sch2:18}, with a very modest dependence on the adopted mass-loss prescription. However, the upper luminosity cut-off in the greater 30\,Dor region is lower than for its central cluster R\,136, implying an upper initial mass limit of $\approx 200\,M_\odot$ compared to $\approx 300\,M_\odot$ in R\,136. This could be an indication of a general dependence of the upper stellar mass limit on the conditions in the star-forming environment.

Finally, the observed number distribution of stars as a function of effective temperature supports the existence of inflated stellar envelopes that are theoretically predicted to form when main-sequence stars reach high L/M ratios. This puts the majority of bright B\,supergiants in the end phase of core hydrogen burning where they display cool temperatures due to their large, inflated radii.

\begin{acknowledgements}
  {\changed GG thanks the Deutsche Forschunsgemeinschaft (DFG) for financial support under grant No. GR 1717/5-1, the Stellar Evolution group at the Argelander Institute for Astronomy in Bonn for sharing their expertise in stellar modelling and providing computational resources, and the anonymous referee for reviewing this work.}
\end{acknowledgements}


\end{document}